\documentclass[aps,a4paper,twocolumn,showpacs,showkeys,preprintnumbers,amsmath,amssymb,10pt]{revtex4-1}

\usepackage{graphicx}
\usepackage{dsfont}
\usepackage{hyperref}
\usepackage{bm}
\usepackage{dcolumn}
\usepackage[mathscr]{eucal}
\usepackage{color}
\usepackage{pst-plot}
\usepackage{amsmath}
\usepackage{comment}
\DeclareMathOperator{\e}{\mathrm{e}}%Euler e
\DeclareMathOperator{\iu}{\mathrm{i}}%imaginary unit
\DeclareMathOperator{\dd}{\mathrm{d}}
\DeclareMathOperator{\id}{\mathds{1}}%identity matrix
\def\be{\begin{equation}}
\def\ee{\end{equation}}

\newcommand{\bra}[1]{\langle #1|}
\newcommand{\ket}[1]{|#1\rangle}

\newcommand{\proj}[1]{\vert #1\rangle\!\langle#1 \vert}
\newcommand{\norm}[1]{\left\| #1 \right\|}

\newtheorem{mydef}{Definition}

\begin{document}
%opening
\title{Trapped surfaces and emergent curved space in the Bose-Hubbard model}
%\textcolor{blue}{Funky hopping of drunk particles on cool graphs (teaser)}}

\author{Francesco Caravelli${}^{1,2,3}$}
\email{fcaravelli@perimeterinstitute.ca}
\author{Alioscia Hamma${}^{1}$}
\email{hamma@perimeterinstitute.ca}
\author{Fotini Markopoulou${}^{1,2,3}$}
\email{fmarkopoulou@perimeterinstitute.ca}
\author{Arnau Riera${}^{4,5}$}
\email{arnauriera@gmail.com}

\affiliation{${}^1$ Perimeter Institute for Theoretical Physics, \\
Waterloo, Ontario N2L 2Y5
Canada, \\and\\
 ${}^2$ University of Waterloo, Waterloo, Ontario N2L 3G1, Canada,\\
 and\\
${}^3$ Max Planck Institute for Gravitational Physics, Albert Einstein Institute,\\
Am M\"uhlenberg 1, Golm, D-14476 Golm, Germany\\
and\\
${}^4$Institute for Physics and Astronomy, University of Potsdam, 14476 Potsdam, Germany\\
and\\
${}^5$Dahlem Center for Complex Quantum Systems, Freie Universit{\"a}t Berlin, 14195 Berlin, Germany
}

%%%%%%%%%%%%%%%%%%%%%%%%%%%%%%%%%%%%%%%%%%%%%%%%%%%%%
%%%%%%%%%%%%%%%%%%%%%%%%%%%%%%%%%%%%%%%%%%%%%%%%%%%%%

\begin{abstract}
A Bose-Hubbard model on a dynamical lattice was introduced in previous work as a spin system analogue of emergent geometry and gravity.  Graphs with regions of high connectivity in the lattice were identified as 
candidate analogues of spacetime geometries that contain trapped surfaces.  
We carry out a detailed study of these systems and show explicitly that the highly connected subgraphs trap matter.  We do 
this by solving the model in the limit of no back-reaction of the matter on the lattice, and for states with certain symmetries 
that are natural for our problem.  We find that in this case the problem reduces to a one-dimensional Hubbard model on a lattice 
with variable vertex degree and multiple edges between the same two vertices.  
In addition, we obtain 
 a (discrete) differential equation for the evolution of the probability density of particles which  is closed in the classical
regime. This is a wave equation in which the vertex degree is related to the local speed of propagation of  probability. This allows
an interpretation  of the probability density of particles similar to that in analogue gravity systems:
matter inside this analogue system sees a curved spacetime. We verify our analytic results by numerical simulations. 
%\ar{\sout{We support our analytic results by a numerical study of the localization and  entanglement entropy of the effective models, for a
%specific set of parameters}}.
Finally, we analyze the dependence of localization on a gradual, rather than abrupt,  fall-off of the vertex degree on the boundary of the highly connected region and find that matter is localized in and around that region.
\end{abstract}

\maketitle
%\tableofcontents

\section{Introduction}
Since the discovery of the Hawking and Unruh effects it has been clear that gravity 
is fundamentally different from the other forces. That a new thermodynamics has to 
be associated to black hole physics is a remarkable puzzle which physicists are slowly unveiling. In the last two decades,  
the possibility that gravity itself may have a thermodynamical origin has been explored from a variety of angles
\cite{MM, ADSCFT, analog, Hor}.
A closely related idea is that gravity may be emergent, either the thermodynamics of a microscopic 
(quantum) theory \cite{Jac, Pad, Ver}, or emergent in the condensed matter sense \cite{Vol, Hu, GuWe, Xu}.

Motivated by the possibility that gravity may be emergent and the questions this raises, {\em quantum graphity} was introduced 
in  \cite{graphity1,graphity2} as a method to study the emergence of geometry and gravity in the simplified but explicit and 
tractable setting of a spin system.  Graphity models are systems of quantum, dynamical graphs where the spin degrees of freedom 
correspond to dynamical adjacency:  the edges of the graphs can be on (connected), off (disconnected), or in a superposition of 
on and off.  By interpreting the graph as pre-geometry (the connectivity of the graph tells us who is neighboring whom), 
this provides a way to treat pre-geometry as a spin system.  A particular graphity model is given by such graph states evolving 
under a local Ising-type Hamiltonian.  The goal is to investigate whether such a local dynamics can lead to the system 
exhibiting aspects of gravity in the thermodynamical limit. 
In previous work we were able to derive features such as a variable speed of light from a non-geometric quantum system; we studied the quantum dynamics in a spin system as a precursor to a quantum theory of gravity ; we showed entanglement and thermalization between geometry and matter;  and found that certain states appear to behave like geometries containing trapped surfaces.

These  models can also be viewed as analogue models for gravity.  In the analogue gravity program various 
phenomena of general relativity (e.g., black holes or cosmological geometries) are modeled by other physical systems, 
such as acoustics in a moving fluid, superfluid helium, or Bose-Einstein condensate; gravity waves in water; 
and propagation of electromagnetic waves in a dielectric medium \cite{analog}.
The fluid analogues provide a very interesting concrete setting in which to study emergent gravitational properties, but leave open several important questions:
 1) What is the role of the background fluid?  
2) Can we go beyond analogues of kinematical only aspects of general relativity?  3) Lorentzian geometry:  in analogue gravity systems, 
Lorentzian geometry ``emerges'' in the sense that it is the geometry seen by excitations inside the fluid (phonons).  What happens if there is no background fluid, and can this work in a quantum system?  

Alternative analogue models, including ours, may attempt to answer some of these.  
For example, in our model there is no background fluid.  There is, however, a background time in the Hamiltonian.  Interestingly, as we will show in this paper,  matter in our model sees an effective Lorentzian geometry, a phenomenon that can 
be traced to the locality of the Hamiltonian.

In this article, we gain substantial insight into  the model of \cite{graphity2} by studying two important questions: we explicitly confirm the argument of \cite{graphity2} that highly connected regions in the graph trap matter, and we relate the evolution of matter under a Hubbard Hamiltonian on these lattices to a wave equation on a curved geometry.  
We achieve this by restricting the model to a large class of states  relevant for these questions.  
These are states in which  the graph has certain symmetries (although the vertex connectivity and the number of edges linking two vertices can vary) and states of  matter that respect these symmetries.  

\begin{figure}
\includegraphics[width=\columnwidth]{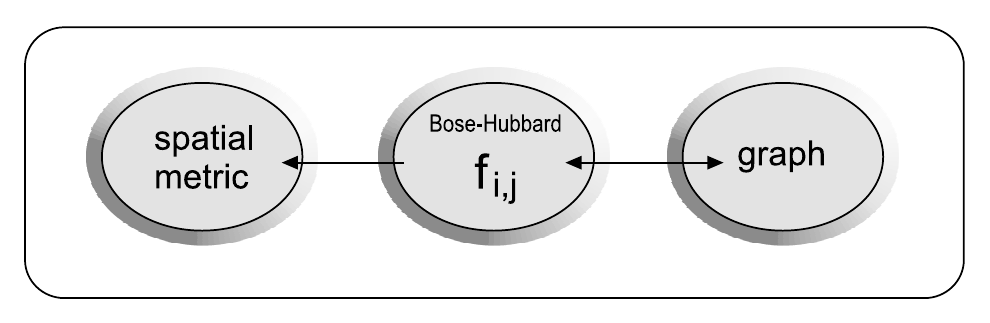}
\caption{The scheme representing the relations between the graph, the hopping energies $f_{i,j}$ of the Bose-Hubbard model and the emergent metric: the graph modifies the strength of the interaction in the Bose-Hubbard Hamiltonian, which in turn translates into a curved geometry (for the appropriate states).}
\label{fig:scheme} 
\end{figure}

We find that the Bose-Hubbard Hamiltonian with homogeneous couplings on a lattice with 
varying vertex degree is equivalent to a Bose-Hubbard Hamiltonian on a regular-degree graph but with site-dependent effective 
couplings (similar to the behavior seen in \cite{Bri}). This makes it possible to connect the coefficients in the Hamiltonian to  geometric properties 
of the graph. The picture which emerges from this analysis is the one of Fig.~\ref{fig:scheme}:
the graph modifies the strength of the interaction in the Bose-Hubbard Hamiltonian, and this appears as a curved geometry to the propagating matter. 

 It is important to note that
the emergent curved space  is a {\em dynamical} property of the system. The geometry that 
the particles propagating on the graph see  depends on the dynamics of the particles and it is not just a property of the 
graph, and, in addition, the resulting motion of the particles will change the graph and so affect the geometry.

The structure of the paper is as follows. In Section \ref{section:model} we review the model introduced in \cite{graphity2} and state the approximations we will use.  This amounts to a Hubbard model on a fixed but unusual lattice. In Section \ref{section:symmetries}, we define foliated and rotationally invariant graphs, the symmetric graph states we will consider, and delocalized states, our similarly symmetric matter states.  In Section \ref{section:trapped}, we study the states that were  previously identified as possible trapped surfaces and confirm this property.  These states have the symmetries defined in the previous Section.  In the single particle sector we can diagonalize the Hamiltonian on the subspace of rotationally invariant states and show that the ground state belongs to that subspace.  We then show that this ground state is protected by a gap which increases linearly with the size of the completely connected subgraph, so when that size is large, this area acts as a trap for matter.  In the many-particle sector, we find that the ground state  is a Bose-Einstein condensate of delocalized particles at the completely connected region, again with a  large gap. We confirm these results numerically.
 In Section \ref{section:EOM},  we relate the one-dimensional Bose-Hubbard model with site-dependent coefficients  to  the
wave equation in curved space for the particle probability on the lattice. In section \ref{section:falloff},  we study numerically  matter localization as a function of a fall-off parameter of vertex degree  on the boundary of the highly connected region.  We find that the particles are localized inside the trapped surface as before, but also on a small region around it. 
Conclusions follow.

\section{The Model}
\label{section:model}

 In  this section we will introduce the model, starting with the basic idea of describing a dynamical lattice using quantum degrees of freedom on the edges of a complete graph.

We start with a quantum mechanical description of a universe of $N_v$ elementary systems, given by the set $\{ \mathscr H_i, \widehat H_i\}$ 
of the Hilbert spaces $\mathscr H_i$ and Hamiltonians $\widehat H_i$ of the systems $i=1,...,N$. This presumes it makes sense 
to talk of the time evolution of an observable with support on $\mathscr H_i$ without any reference to a spatial coordinate for $i$.

We choose $\mathscr H_i$ to be the Hilbert space of a harmonic oscillator. We denote its creation and destruction 
operators by $b^\dagger_i$ and $b_i$ respectively, satisfying the usual bosonic commutators. 
Our $N_v$ physical systems then 
are $N_v$ bosonic modes and the total Hilbert space of such modes is given by
\begin{equation}
\mathscr H_{bosons} = \bigotimes_{i=1}^{N_v} \mathscr H_i.
\end{equation}
If the harmonic oscillators are not interacting, the total Hamiltonian is trivial:
\begin{equation}
\label{hv}
\widehat H_v = \sum_{i=1}^{N_v} \widehat H_i =- \sum_{i=1}^{N_v} \mu b^\dagger_i b_i.
\end{equation}
If, instead, the harmonic oscillators are interacting, we need to specify who is interacting with whom.  Let us 
call $ I$ the set of the pairs of oscillators ${\bf e}\equiv(i,j)$ that are interacting. Then the Hamiltonian  
reads as
\begin{equation}
\widehat H = \sum_i \widehat H_i + \sum _{{\bf e}\in  I} \widehat h_{\bf e},
\label{eq:H}
\end{equation}
where $\widehat h_{\bf e}$ is a Hermitian operator on $ H_i \otimes  H_j$ representing the interaction between the system $i$ 
and the system $j$.

Spacetime geometry determines   the set of relations among physical systems, namely, interactions are only possible between neighbors.  
 In a discrete setup,
 we can introduce a primitive notion of spacetime geometry via the adjacency
matrix $A$, the $N_v\times N_v$ symmetric matrix defined as
\begin{equation}
A_{ij}=\left\{{\begin{array}{ll}
1&{\mbox{if $i$ and $j$ are adjacent}}\\
0&{\mbox{otherwise}}.
\end{array}}
\right.
\end{equation}
$A$ defines a graph on $N_v$ nodes, with an edge between nodes $i$ and $j$ for every $1$ entry in the matrix.  
Now note that all possible graphs on $N_v$ vertices are subgraphs of $K_N$, the 
complete graph on $N_v$ vertices (with adjacency matrix $A_{ij}=1$ for every $i,j$).  In our model, the $N_v(N_v-1)/2$ edges of $K_N$ correspond to (unordered) pairs 
${\bf e}\equiv(i,j)$ of harmonic oscillators.  To every such pair ${\bf e}$ (an edge of $K_N$)  we 
associate a Hilbert space $\mathscr H_{\bf e}\simeq {\bf C}^2$ of a spin $1/2$.
The total Hilbert space for the graph edges is then
\begin{equation}
\mathscr H_{graph} = \bigotimes_{\bf e =1}^{N_v(N_v-1)/2} \mathscr H_{\bf e}.
\end{equation}

The basis in $\mathscr H_{graph}$ is chosen so that to every subgraph $G$ of $K_N$ corresponds a basis element 
in $\mathscr H_{graph}$: the basis element $| e_1\ldots  e_{N_v(N_v-1)/2}\rangle\equiv|G\rangle$ corresponds to the 
graph $G$ that has an edge ${\bf e}_s$ present for every $e_s =1$. For every edge $(i,j)$, the corresponding $SU(2)$ 
generators will be denoted as $S^i = 1/2 \sigma^i$ where $\sigma^i$ are the Pauli matrices.

The total Hilbert space of the theory is
\begin{equation}
{\mathscr H} = \mathscr H_{bosons}\otimes \mathscr H_{graph},
\end{equation}
and therefore a basis state in $ H$ has the form
\begin{eqnarray}
|\Psi\rangle &\equiv& |\Psi^{(bosons)}\rangle\otimes|\Psi^{(graph)}\rangle \\
 &\equiv& |n_1,...,n_{N_v}\rangle\otimes |e_1,...,e_{\frac{N_v(N_v-1)}{2}}\rangle.
\end{eqnarray}
The first factor tells us how many bosons there are at every site $i$.
 The second factor tells us which pairs $(i,j)$ interact, that is,  
{\em the structure of interactions is now promoted to a quantum degree of freedom}. 
Unlike the model presented in \cite{graphity2} where {\em hard core bosons} were considered, 
the interaction  among the bosons in the present paper  is an on-site potential described by
$u b^\dagger_i b_i(b^\dagger_i b_i-1)/2$ where $u$ is the interaction energy between two particles.

In general, a quantum state in the full model describes a system in a generic superposition 
of energies of the harmonic oscillators, and of interaction terms among them. A state can be a quantum 
superposition of ``interactions''. For instance, consider the systems $i$ and $j$ in the state
\begin{equation}\label{s1}
|\phi_{ij}\rangle = \frac{|10\rangle\otimes|1\rangle_{ij}+ |10\rangle\otimes|0\rangle_{ij}}{\sqrt{2}}.
\end{equation}
This state describes the system in which there is a particle in $i$ and no particle in $j$, and 
a superposition of link and no link between $i$ and $j$. 

A simple but interesting matter interaction term is the one that describes the physical process in which a quantum in the 
oscillator $i$ is destroyed and one in the oscillator $j$ is created. This dynamical process is possible when
there is an edge between $i$ and $j$. Such dynamics is described by a Hamiltonian of the form
\begin{equation}\label{hhop}
\widehat H_{\textrm{hop}} = -E_{\textrm{hop}}\sum_{(i,j)}\widehat P_{ij}\otimes  (b^\dagger_i b_j +b_i b^\dagger_j),
\end{equation}
where
\begin{equation}
\widehat P_{ij}\equiv \widehat S^+_{(i,j)} \widehat S^-_{(i,j)} =|1\rangle\langle 1|_{(i,j)} = \left( \frac{1}{2}-\widehat S^z \right)_{(i,j)}
\end{equation}
is the projector on the state such that the edge $(i,j)$ is present and the spin operators are defined 
as $\widehat S^+_{(i,j)} = |1\rangle\langle0|_{(i,j)}$ and $\widehat S^-_{(i,j)} = |0\rangle\langle1|_{(i,j)}$. With this Hamiltonian, 
the state {defined in} Eq.(\ref{s1}) can be interpreted as the quantum superposition of a particle that may hop or not from 
one site to another. 

It is possible to design such systems in the laboratory. For instance, one can use arrays 
of Josephson junctions whose interaction is mediated by a quantum dot with two levels.

We note that it is the
dynamics of the particles described by $\widehat H_{\textrm{hop}}$ that gives to the degree of freedom $|e\rangle$ the meaning of 
geometry.
The geometry at a given instance is given by the set of relations describing the dynamical potentiality for hopping.
 Two vertices $j$, $k$ can be ``empty'', that is, the oscillators $j,k$ are in the ground state, but they can still  serve to allow a particle to
hop from $i$ to $j$, then to $k$, then to $l$. {\em We read out the structure of the graph from the interactions, not 
from the mutual positions of particles.}

In addition, $\widehat H_{\textrm{hop}}$   tells us that it takes a finite amount of time to go from $i$ to $j$. For instance, in a one-dimensional graph (a chain) it takes a finite amount of time  for a particle to go from one end 
of the chain to another (modulo exponential decaying terms). As shown in \cite{HMPSS}, this results in a ``spacetime'' picture  with 
a finite light-cone structure. 

The hopping amplitude is given by $t$, and therefore all the bosons have the same speed. 
Note that, for a larger Hilbert space on the links, we can have different speeds for 
the bosons. 

Of course, we need a Hamiltonian also for the spatial degrees of freedom alone. The simplest choice is 
to assign an energy to every edge:
\begin{equation}\label{hlink}
 \widehat H_{link} = -U\sum_{(i,j)} \sigma^z_{(i,j)}.
\end{equation}

\begin{figure}
\includegraphics[width=\columnwidth]{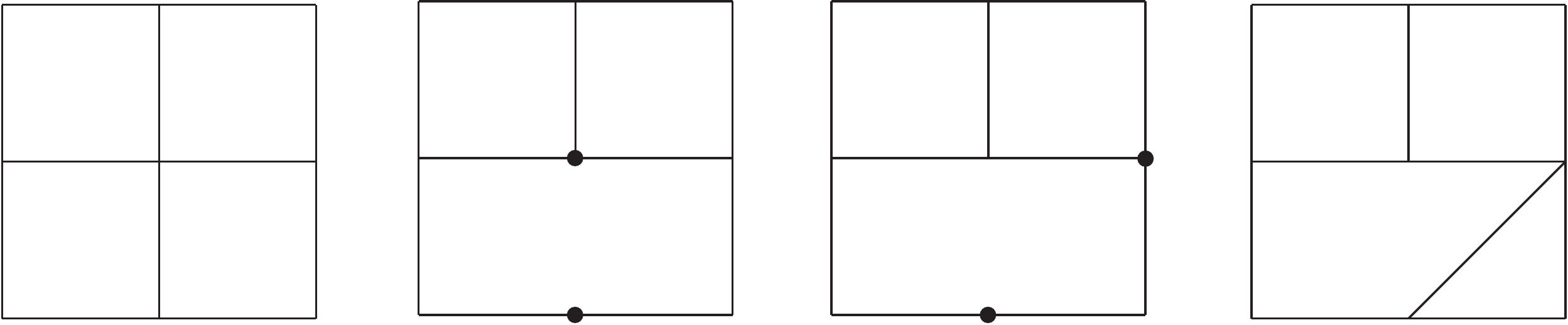}
\caption{Diagramatic representation of the exchange and hopping terms of the model.  The exchange term destroys a link and creates two particles on the vertices of that link; the particles hop; two particles can combine to form a new link. }
\label{fig:FullModel} 
\end{figure}

Finally, a central motivation for the model \cite{graphity2} was to have space and matter interact unitarily. The term
\begin{equation}
\widehat H_{ex} =k \sum_{(i,j)}  \left( \widehat S^-_{(i,j)}\otimes ( b^\dagger_i
b^\dagger_j)^Q +\widehat S^+_{(i,j)}\otimes (b_i b_j)^Q\right),
\end{equation}
introduced in \cite{graphity2},  destroys an edge $(i,j)$ and creates $Q$ quanta at $i$ and $Q$ quanta at $j$, or, vice-versa, destroys $Q$ quanta 
at $i$ and $Q$ quanta at $j$ to convert them into an edge (see Fig.\ \ref{fig:FullModel}).

As mentioned above, the long-term ambition of these models is to find a quantum Hamiltonian that is a spin system analogue of gravity.  
In this spirit, matter-geometry interaction is desirable as it is a central feature of general relativity.  The above dynamics can be considered as a very simple 
first step in that direction.  

This was the final step that brings us to the total Hamiltonian of the model which is
\begin{equation}\label{h}
\widehat H = \widehat H_{link}+ \widehat H_v + \widehat H_{ex} + \widehat H_{\textrm{hop}}.
\end{equation}
This is the original model proposed in \cite{graphity2}. It is a Bose-Hubbard model on a dynamical lattice, where the projector on the modified hopping term means that the dynamics gives the lattice the meaning of space:  space is where matter is allowed to go.

In the present work, we study the model for a particular class of graphs that have been conjectured to be analogues of trapped surfaces.   
We are interested in the approximation $k\ll t$,
which can be seen as the equivalent of ignoring the backreaction of the matter on the geometry.  For simplicity, we will 
set  $U=k=0$, meaning that
\begin{equation}
 \widehat H = \widehat H_v  + \widehat H_{\textrm{hop}}.
\label{eq:redH}
\end{equation}
In this case, 
the total number of particles on the graph is a conserved charge.  $\widehat H_v$ and $\widehat H_{links}$ are constants on 
fixed graphs with fixed number of particles.   The  Hamiltonian is the ordinary Bose-Hubbard model
on a fixed graph, but that graph can be unusual, with sites of varying connectivity and with more than one edge connecting two sites.
Our aim will then be to study the effective geometry that matter in this model 
sees.  

Even on a fixed lattice, the Hubbard model is difficult to analyze, with few results in higher dimensions.  It would seem that our problem, propagation on a lattice with connectivity which varies from site to site is also very difficult.  Fortunately, it turns out that for our purposes it is sufficient to restrict attention to lattices with certain symmetries.

%%%%%%%%%%%%%%%%%%%%%%%%%%%%%%%%%%%%%%%%%%%%%%%%%%%
\section{Foliated and rotationally invariant graphs}
\label{section:symmetries}

In this Section we define graphs with two particular properties which we call {\em foliation} and {\em rotational invariance}.  These properties will allow us to greatly simplify 
the  calculations that follow without loss of generality.
We will see that the problem of finding the ground state of hopping Hamiltonians,  on graphs with these properties
can be simplified to the solution of the one-dimensional Bose-Hubbard model.

\subsection{Foliated graphs}

\begin{figure}
\includegraphics[width=\columnwidth]{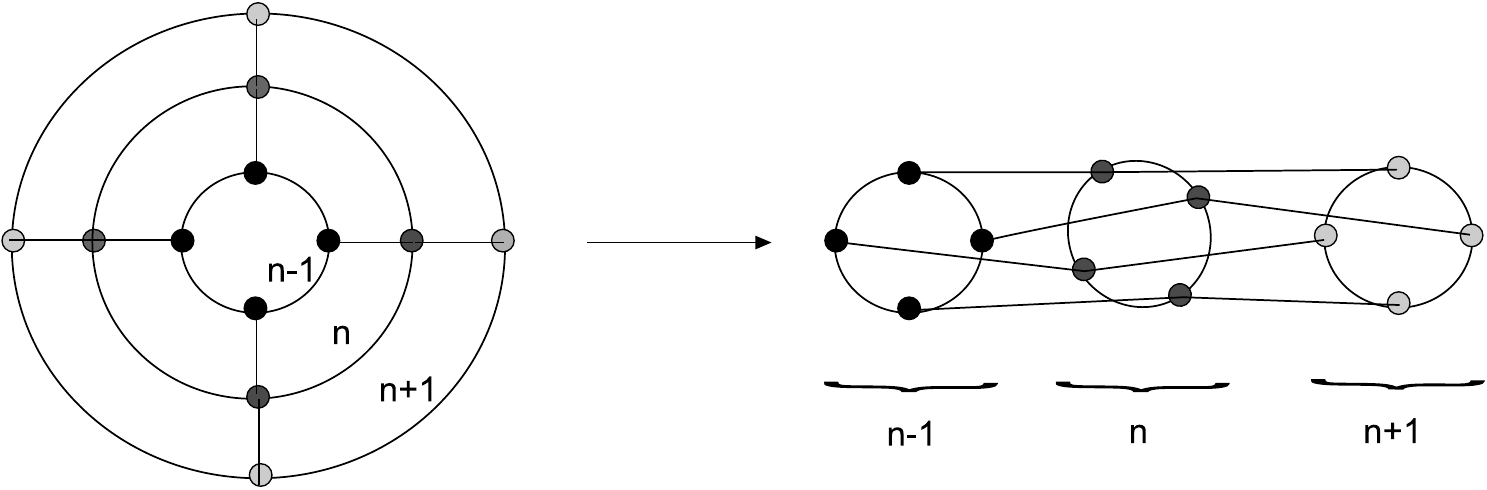}
\caption{A 2-d lattice  that can be foliated.}
\label{fig:2d}
\end{figure}

\begin{figure}
\includegraphics[width=\columnwidth]{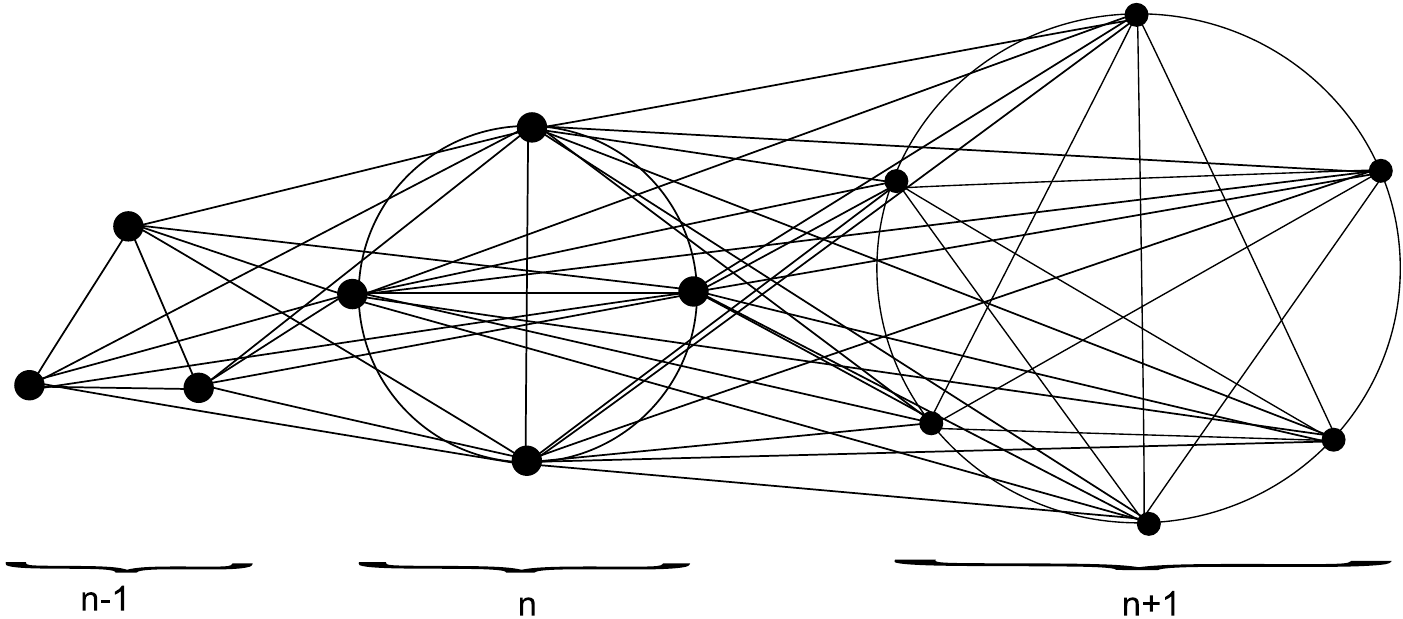}
\caption{A non-planar lattice that can be 1-foliated.}
\label{fig:2d_nonplan}
\end{figure}

A \emph{foliated graph} is a graph that can be decomposed into a set of subgraphs connected by edges in a successive way. More precisely,
let $g_i$ be a labeling of subgraphs of a graph $G$ and $E_{i}$ a labeling of the set of edges connecting the sets $g_i$, such that
 $\cup_i (g_i \cup E_i) = G$. Then,

\begin{mydef}
A graph $G$ is \emph{foliated} if it can be decomposed in several disjoint subgraphs $g_i$ with the 
following properties:
\begin{enumerate}
\item All the subgraphs $g_i$ are degree regular.
\item All the edges of a subgraph $g_i$ connect a vertex in $g_i$ to a vertex in   $g_{i-1}$ or $g_{i+1}$.
\item The number of edges connecting a vertex in  $g_i$ ($g_{i+1}$) to vertices in $g_{i+1}$ ($g_i$) is the same for every vertex of $g_i$ ($g_{i+1}$). 
      This number is called \emph{relative degree} and is represented by $d_{i,i+1}$ ($d_{i+1,i}$).
\end{enumerate}
\end{mydef}
Notice that the name, \emph{foliated}, comes from the fact that these graphs can be decomposed into subgraphs such as
any foliated structure can be separated into thin layers.
Examples of foliated graphs are presented in Figs.~\ref{fig:2d} and  \ref{fig:2d_nonplan}.

The number of edges that connect two consecutive subgraphs $g_i$ and $g_{i+1}$ is given by
\be
d_{i,i+1} N_i = d_{i+1,i}N_{i+1} \, ,
\label{graphpro}
\ee
where $N_i$ is the number of vertices of $g_i$.

%%%%%%%%%%%%%%%%%%%%%%%%%%%%%%%%%%%%%%%%%%%%%%%%
%\subsection{Particle propagation on foliated graphs}

In order to discuss the properties of the hopping Hamiltonians defined by foliated graphs, 
let us introduce a natural coordinates $(i,m)$ to specify a vertex of a foliated graph.
The coordinate $i$ specifies the subgraph $g_i$ and $m$ refers to the particular vertex in $g_i$.
The hopping Hamiltonian of a foliated graph in terms of these coordinates can be written as
\begin{align}
\widehat H_{\textrm{fol}}&=-\sum_{i=0}^{R-1}\sum_{m,m'=0}^{N_i-1} A_{mm'}^{(i)} b_{i,m}^\dagger b_{i,m'}+h.c.\nonumber \\
&-\sum_{i=0}^{R-1}\sum_{m=0}^{N_i-1}\sum_{m'=0}^{N_{i+1}-1} B_{mm'}^{(i)}b_{i,m}^\dagger b_{i+1,m'}+h.c.\, ,
\label{eq:hopping-ham-foliated}
\end{align}
where $b_{i,m}^\dagger$($b_{i,m}$) is the creation (annihilation) operator of a particle at the vertex $(i,m)$,
$R$ is the number of layers $g_i$, $A_{mm'}^{(i)}$ is the adjacency matrix of the subgraph $g_i$, and
$B_{mm'}^{(i)}$ stands for the edges between the vertices of $g_i$ and $g_{i+1}$.

In this model, the \emph{delocalized} states on the subgraphs $g_i$ are particularly interesting.  These are  states of a particle which is completely and uniformly spread over the graph $g_i$, defined as follows:

\begin{mydef}
The \emph{delocalized} state, $\ket{i}$, in $g_i$ is defined by
\be
\ket{i}=\frac{1}{\sqrt{N_i}}\sum_{m=0}^{N_i-1} \ket{i,m}.
\ee
\end{mydef}

The delocalized state $\ket{i}$
is an eigenstate of the hopping Hamiltonian defined from the degree-regular subgraph $g_i$,
\be
\widehat H_{\textrm{fol}}^{(i)}=-\sum_{m,m'=0}^{N_i-1} A_{mm'}^{(i)} b_{i,m}^\dagger b_{i,m'} +h.c.\, .
\ee
More explicitly,
\be
\widehat  H_{\textrm{fol}}^{(i)} \ket{i}=\frac{1}{\sqrt{N_i}} \sum_{m=0}^{N_i-1} \left(\sum_{m'=0}^{N_i-1}A_{m'm}^{(i)}\right)\ket{i,m}= - d_{i,i} E_{hop} \ket{i}\ ,
\ee
where we have used the degree-regularity of the graph, $d_{i,i}=\sum_{m}A_{nm}^{(i)}$, for all $n$.
For example, when the graph $g_i$ is a chain, the completely delocalized state is the ground state of 
the system and has energy $-2$, with $2$ is the degree of the vertices of a 1-dimensional  chain.

The subspace spanned by the delocalized states $\ket{i}$ is an eigenspace of the system,
since the projector onto this subspace,
\be
 \widehat P_{\textrm{d}} = \sum_{i=0}^{R-1} \proj{i}\, ,
\label{eq:proj}
\ee
commutes with the Hamiltonian \eqref{eq:hopping-ham-foliated}, $[H_{\textrm{fol}},P_{\textrm{d}}]=0$. 
Therefore,
the time evolution of any superposition of delocalized states lies always in this subspace (it is a superposition of delocalized states).
This allows us to define an effective Hamiltonian for the delocalized states,
\begin{align}
\label{eq:final}
 \widehat H_{\textrm{eff}} &:= \widehat P_{\textrm{d}} \widehat H_{\textrm{fol}} \widehat P_{\textrm{d}} \\
       &= \sum_{i=0}^{R-1} \left( f_{i,i+1} \left( \ket{i+1}\bra{i}+\ket{i}\bra{i+1}\right)+{\mu}_{i} |i\rangle\langle i|\right) ,\nonumber
\end{align}
where $\mu_i=\bra{i}\widehat H_{\textrm{fol}}\ket{i}=-E_{\textrm{hop}}d_{i,i}$  and 
\begin{equation}
 f_{i,i+1}:= \bra{i}\widehat H_{\textrm{fol}}\ket{i+1}=-E_{\textrm{hop}}\ d_{i,i+1} \sqrt{\frac{ N_{i}}{N_{i+1}}}\, .
\end{equation}
Note that the Hamiltonian \eqref{eq:final} is a one-dimensional Bose-Hubbard Hamiltonian with chemical potential $\mu_i$ and tunneling coefficient $f_{i,i+1}$.

The mass term (or chemical potential) $\mu$,  is \emph{fattened by  the edges connecting
the nodes within the subgraph $g_i$ of the foliated graph}. This is one the main results of the paper. This behavior
resembles a scalar field not-minimally coupled to (classical) gravity, where the mass of the particle is multiplied by a curvature factor.
In our case, the role of the curvature is played by $-E_{hop} d_{i,i}$ (see Sec.~\ref{section:EOM}). 

The extension
to higher dimensions  is straightforward.  It requires the extension of one-dimensional foliated graphs to graphs which can be foliated in multiple directions,
thus resembling an ordinary lattice, but with multiple links between pairs of sites. The coefficients $f_{k,k-1}$ will depend on the direction of the foliation that the particle
is hopping to. 

We have found an eigenspace of the foliated Hamiltonians for which the effective Hamiltonian is particularly simple.  However, 
 this eigenspace will not, in general, contain relevant eigenstates of the Hamiltonian such as the ground state.
In order to ensure this, we require another symmetry.

\subsection{Rotationally invariant graphs}
Let us present next our definition of the rotationally invariant graphs.
\begin{mydef}
A graph $G$ is called $N$-\emph{rotationally invariant} if there exists an embedding of $G$ to the plane that is invariant
by rotations of an angle $2\pi / N$.
\end{mydef}

With the exception of those graphs that have a vertex in the center, 
$N$-\emph{rotationally invariant} graphs allow for a labeling of their vertices $(n,\theta)\in \mathbb{N}\times \mathbb{Z}_N$
such that their adjacency matrix $A_{(n\theta) \, , (n'\theta')}$ only depends on $n$, $n'$ and $\theta-\theta'$.
We can make use of these coordinates $(n, \theta)$ in order to write
the Hamiltonian defined by a rotationally invariant graph
\begin{align}
H_{\textrm{rot}}&=-\sum_{\theta=0}^{N-1}\sum_{n,n'} A_{nn'}b_{n\theta}^\dagger b_{n'\theta} +h.c.\nonumber \\
&-\sum_{\theta=0}^{N-1}\sum_{\varphi=1}^{N-1}\sum_{n,n'} B_{n,n'}^{(\varphi)}b_{n\theta}^\dagger b_{n' \theta+\varphi}+h.c., 
\end{align}
where $b_{n,\theta}^\dagger$ ($b_{n,\theta}$) is the creation (annihilation) operator at the vertex $(n,\theta)$,
$A_{nn'}$ is the adjacency matrix of the graph of any angular sector and $B_{n,n'}^{(\varphi)}$
is the adjacency matrix of two angular sectors at an angular distance $\varphi$ in units of $2\pi / N$.
 
Let us introduce the rotation operator $\widehat L$ defined by
\begin{align}
\widehat L b_{n,\theta} = b_{r,\theta +1} \widehat L \, \nonumber \\
\widehat L b_{n,\theta}^\dagger = b_{r,\theta +1}^\dagger \widehat L \, .
\label{eq:L}
\end{align}
The effect of the operator $L$ is particularly easy to understand in the single particle case:
\be
\widehat L \ket{n,\theta} = \widehat L b_{n,\theta}^\dagger \ket{0} = b_{n,\theta +1}^\dagger \widehat L \ket{0}  = \ket{n,\theta+1}\, ,
\ee
where we have assumed that the vacuum is invariant under a rotation $\widehat L\ket{0}=\ket{0}$.

Note that $\widehat L$ is unitary 
and its application $N$ times gives the identity, $\widehat L^N=\id$. 
This implies that its eigenvalues are integer multiples of $2\pi / N$.

Another interesting property of $L$ is that commutes both with the rotationally invariant 
Hamiltonians and with the number operator $\widehat N_p$,
\be
	[\widehat H_{\textrm{rot}},\widehat L]=[\widehat N_p,\widehat L]=[\widehat H_{\textrm{rot}},\widehat N_p]=0 \, .
\ee
Therefore $\widehat H_\textrm{rot}$, $\widehat N_p$, and $\widehat L$ form a complete set of commuting observables and
the Hamiltonian is diagonal in blocks of constant $\widehat L$ and $\widehat N_p$.
This allows us to simplify the problem of finding the ground state and the first excited states of the system, 
a very useful fact we will use in the next section. 

\begin{figure}
 \includegraphics[scale=0.5]{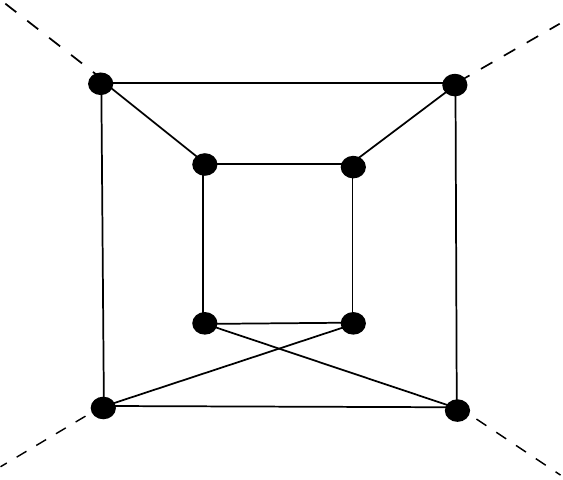}
\caption{A foliated graph which is not rotationally invariant.}
\label{fig:counterex}
\end{figure}

Note that rotational invariance and foliability are not equivalent. 
There are graphs which can be foliated and are not rotationally invariant and vice-versa.
An example is given in Fig. \ref{fig:counterex}.

We have seen that both for the cases of foliated graphs and the rotationally invariant graphs, 
{\em the subspace spanned by the completely delocalized states is an eigenspace of the system. }
In the foliated model, effective dynamics in this eigenspace are given by a one-dimensional Bose-Hubbard Hamiltonian. 
In the rotational invariant case, this eigenspace contains the ground state of the system.
Thus, the ground state of graphs with both symmetries lies in the subspace of completely delocalized states
and the computational effort to find it is equivalent to the solution of a one-dimensional Bose-Hubbard model.
This allows us to analyze a complicated model using the approximation of a one-dimensional spin chain, with obvious advantages, especially for numerical work.

%%%%%%%%%%%%%%%%%%%%%%%%%%%%%%%%%%%%%%%%%%%%%%%%

\section{Regions of high connectivity as trapped surfaces}
\label{section:trapped}

\begin{figure}
\includegraphics[scale=0.7]{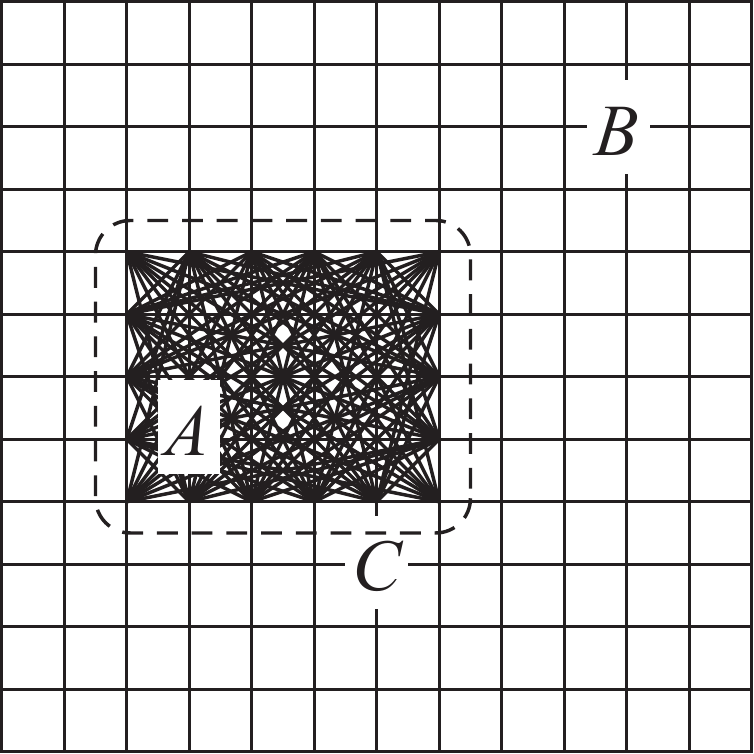} 
\caption{A region of higher connectivity in a regular graph.}
\label{fig:densit}
\end{figure}

In  \cite{graphity2}, we observed that the Hamiltonian (\ref{eq:redH}) caused 
trapping of matter in regions of higher connectivity. 
The basic mechanism is the following: consider a graph consisting of two set of nodes, $A$ and $B$, separated by a set of points $C$ on the 
boundary. Let the vertices in $A$ be of much higher degree than the vertices in $B$, $d_A\gg d_B$  (see Fig.\ \ref{fig:densit}). If the number of edges departing from the set $C$ and going to the set
$A$ is much higher than the number of edges going from $C$ to $B$, then the hopping particles have a high probability 
of being ``trapped'' into the region $A$.

\begin{figure}
\includegraphics[scale=0.7]{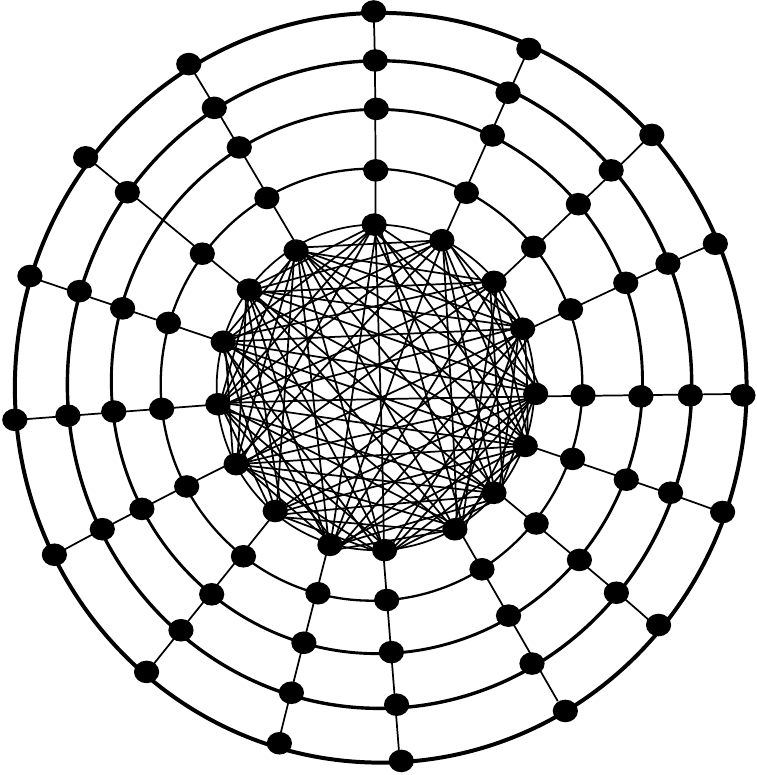}
\caption{The $\mathscr K_N$ graph. }
\label{fig:2dkn}
\end{figure}

Our task in this paper is to make this heuristic argument precise and determine whether these 
high connectivity configurations are spin-system analogues of trapped surfaces.
We will do this by studying specific states that are graphs with symmetries that contain a core 
(trap) of $N$ nodes.  Fig.~\ref{fig:2dkn} is an example of such a graph. Such states
are 1-foliated graphs and we will be able to use the properties we discovered above.

\subsection{Classical trapping}
In order to gain some intuition on the trapping, let us consider the classical analogue of the problem. 
In the classical setup, a particle has a well-defined position in some site of $A$. 
At each time step of length $\hbar / E_{\textrm{hop}}$, the particle randomly jumps to another site of the graph
connected to its current site by an edge.
This process is successively repeated until the particle escapes from $A$.
How much time is required (on average) for  the particle to escape from
the highly connected region?

In this simple model, the probability that the particle jumps to a site outside $A$ is 
\be
p_{\textrm{esc}}=\frac{N_{\textrm{ext}}}{N-1+N_{\textrm{ext}}}\, ,
\ee
where $N_{\textrm{ext}}$ is the number of links that connect a site in $A$ with the environment, and $N$ is
the total number of sites in $A$.
The evaporation time is then given by
\be
t_{\textrm{ev}}= \frac{\hbar}{E_{\textrm{hop}}}\frac{1}{p_{\textrm{esc}}}\approx \frac{\hbar}{E_{\textrm{hop}}}\frac{N}{N_{\textrm{ext}}}\, ,
\ee
where we have assumed that $N \gg N_{\textrm{ext}}$.
In the large $N$ limit, the particle is trapped in $A$.

\subsection{Quantum case: the $\mathscr K_N$ configuration}
In this section we study the spectrum and the configuration of the ground state and the 
finite temperature states of the hopping Hamiltonian on the graph of Figure \ref{fig:2dkn} which we will call the $\mathscr K_N$ graph.  This is a 1-foliated graph with a completely connected core.  
We will show that the model on $\mathscr K_N$ can be solved analytically in the thermodynamic limit.

The position of a vertex of the $\mathscr K_N$ graph can be specified
by means of the integer spherical coordinates $r$ and $\theta$, 
with ranges $0\le r \le R-1$ and $0 \le \theta \le N-1$. 
Then, the quantum state of a particle with a well-defined position in the graph can be written as
\be
\ket{r, \theta} = b_{r, \theta}^{\dagger} \ket{0} \, ,
\ee
where $\ket{0}$ is the vacuum state and $b_{r, \theta}^\dagger$ the corresponding creation operator.
Using these coordinates, the hopping Hamiltonian defined by the $\mathscr K_N$ graph becomes
\begin{align}
\label{eq:Hamiltonian-free-Kn}
\widehat H_0&=\sum_{\theta=0}^{N-1}\sum_{r=0}^{R-1}\left(b_{r+1,\theta}^\dagger b_{r,\theta} + b_{r,\theta + 1}^\dagger b_{r,\theta} + \textrm{h.c.} \right)\\
  &\qquad+ \sum_{|\theta-\theta'|\ge 2} b_{0,\theta'}^\dagger b_{0,\theta}\, , \nonumber
\end{align}
where h.c.\ is the Hermitian conjugate. 
Note that the second term in Eq.~\eqref{eq:Hamiltonian-free-Kn} corresponds to  
hopping in the completely connected region, while the first sum is hopping outside that core. 
Our question is how the introduction of this completely connected region (the second sum)
changes the spectrum and the eigenstates of the system.

\emph{Single particle case}. 
Let us first work out the single particle sector of the Hamiltonian.

In order to determine the eigenstates and eigenvalues of $\widehat H_0$, 
we will
 write the Hamiltonian in the eigenbasis of the rotation operator $\widehat L$ defined in equation (\ref{eq:L}). 
The eigenstates of $\widehat L$ read
\be
\ket{r,\ell}=\frac{1}{\sqrt{N}}\sum_{\theta=0}^{N-1} \e^{\iu \frac{2\pi}{N}\ell \theta}\ket{r,\theta}\, ,
\ee
with eigenvalues
\be
\widehat L \ket{r,\ell} = \e^{\iu \frac{2\pi}{N}\ell}  \ket{r,\ell}\, ,
\ee
where $\ell=0,1,\ldots,N-1$.
The Hamiltonian is diagonal in blocks of constant $\ell$ and 
can be written as
\be
\widehat H = \sum_{\ell=0}^{N-1} \widehat H_\ell\, ,
\ee
where $\widehat H_\ell= \widehat P_\ell \widehat H \widehat P_\ell$ are the projections
onto the eigenspaces of $ \widehat L$ with the projectors $\widehat P_\ell$  defined as
\be
\label{eq:projectorsR}
\widehat P_{\ell}=\sum_{r=0}^{R-1} \proj{r,\ell} \, .
\ee
Inserting Eq.~\eqref{eq:projectorsR} into the definition of $\widehat H_\ell$, we get
\begin{align}
\label{eq:H_ell}
\widehat H_\ell&= -N \delta_{\ell 0} \proj{0}- 2 \cos\left(\frac{2\pi}{N}\ell \right)\theta\left(r -\frac{1}{2}\right)\proj{r} \nonumber \\
&\qquad - (\ket{r+1}\bra{r}+\ket{r}\bra{r+1})\, ,
\end{align}
where $\theta(\cdot)$ is the Heaviside step function, introduced to make the second term of the
right hand side vanish when $r=0$. 

Note that $ \widehat H_\ell$ is  translationally invariant in the limit $R\rightarrow\infty$ 
 for $\ell >0$.
Therefore, it can be analytically diagonalized by the discrete Fourier transform
\be
\label{eq:radial-FT}
\ket{k_r, \ell}=\frac{1}{\sqrt{R}}\sum_{r=0}^{R-1}\e^{\iu \frac{2\pi}{R}k_r r}\ket{r, \ell}\, ,
\ee 
with $k_r=0,\ldots,R-1$ and $1 \le \ell \le N-1$.
We find
\be
\label{eq:eigenproblem-radial}
\widehat H_\ell \ket{k_r, \ell} = 2\left(1-\cos \left(\frac{2\pi}{R}k_r\right)- \cos\left(\frac{2\pi}{N}\ell \right) \right)\ket{k_r, \ell}\, 
\ee
up to $1/R$ corrections that vanish in the infinite limit.

Let us now consider the subspace $\ell=0$ of the rotationally invariant states.
Because the Hamiltonian commutes with $\widehat L$, the ground state $\ket{GS}$  of the system must  
be invariant under its action: $\widehat L \ket{GS} = \ket{GS}$. Therefore, $\ket{GS}$ belongs to this subspace.
On it, we can explicitly construct the matrix for the Hamiltonian $\widehat H_{\ell=0}$:
\begin{equation}
\widehat H_{\ell =0}=-E_{\textrm{hop}}\left( \begin{array}{ccccc}
N & 1 & 0 & \cdots & 0 \\
1 & 2 & 1 & \ddots & \vdots \\
0 & 1 & 2 & \ddots & 0 \\
\vdots &  \ddots & \ddots & \ddots & 1 \\ 
0 & \cdots & 0 & 1 & 2 \\
\end{array} \right).
\label{eq:effmatrix}
\end{equation}
It is a tridiagonal matrix with
 characteristic polynomial  $p_R(\lambda)=\det(\widehat H-\lambda I)$ which
can be written in a recursive way as
\begin{align}
p_0(\lambda)&=1,\nonumber \\
p_1(\lambda)&=-N \, , \\
p_n(\lambda)&=-(2-\lambda)p_{n-1}(\lambda)- p_{n-2}(\lambda) \nonumber \, .
\end{align}
Note that because of the recursive relation, 
it is not clear whether the other eigenvalues apart from the first one depend on $N$ or not. 

It is easy to see that, if we rescale $E_{\textrm{hop}}=\tilde E_{\textrm{hop}}/N$ and 
take $N\rightarrow \infty$, the only element left in the matrix is the element  associated to the
$|0\rangle$ state. 
%This means that the only eigenvalue that depends on $N$ 
%is that of the $|0\rangle$ state because $p_1(\lambda)$ is already an eigenvalue.
Thus, at $N\rightarrow\infty$, the ground state becomes $\ket{r=0,\ell=0}$ and 
the gap between it and the first excited state 
scales as $N$.  
In the thermodynamic limit, the ground state of the single particle sector corresponds to a particle completely delocalized
in the complete graph,
\be
\lim_{N\to \infty} \ket{GS} = \ket{r=0, \ell=0} = \frac{1}{\sqrt{N}}\sum_{\theta=0}^{N-1}\ket{0,\theta}\, .
\ee

The rest of eigenvectors of the subspace $\ell=0$ are orthogonal to $\ket{GS}=\ket{r=0,\ell=0}$, and therefore  lie in
the subspace spanned by $\ket{r, \ell=0}$ with $r\ge 1$. The Hamiltonian can be analytically diagonalized in this subspace
by the same transformation used in Eq.~\eqref{eq:radial-FT}:
\be
\label{eq:radial-FT-l=0}
\ket{k, 0}=\frac{1}{\sqrt{R}}\sum_{r=1}^{R-1}\e^{\iu \frac{2\pi}{R}k r}\ket{r, 0}\, ,
\ee 
with $k=0, \ldots, R-2$. 

In conclusion, we have seen that {\em the $\mathscr K_N$ model has a unique ground state which is protected by a gap which increases 
linearly with the size $N$ of the completely connected region}. The rest of the eigenvalues form an energy band which
is the same as if we had the $\mathscr K_N$ graph without the completely connected region,
\be
E_{k \ell}= 2\left(1-\cos \left(\frac{2\pi}{R}k_r\right)- \cos\left(\frac{2\pi}{N}\ell \right)\right)\, .
\label{eq:changeinchem}
\ee

\begin{figure}
\includegraphics[width=\columnwidth]{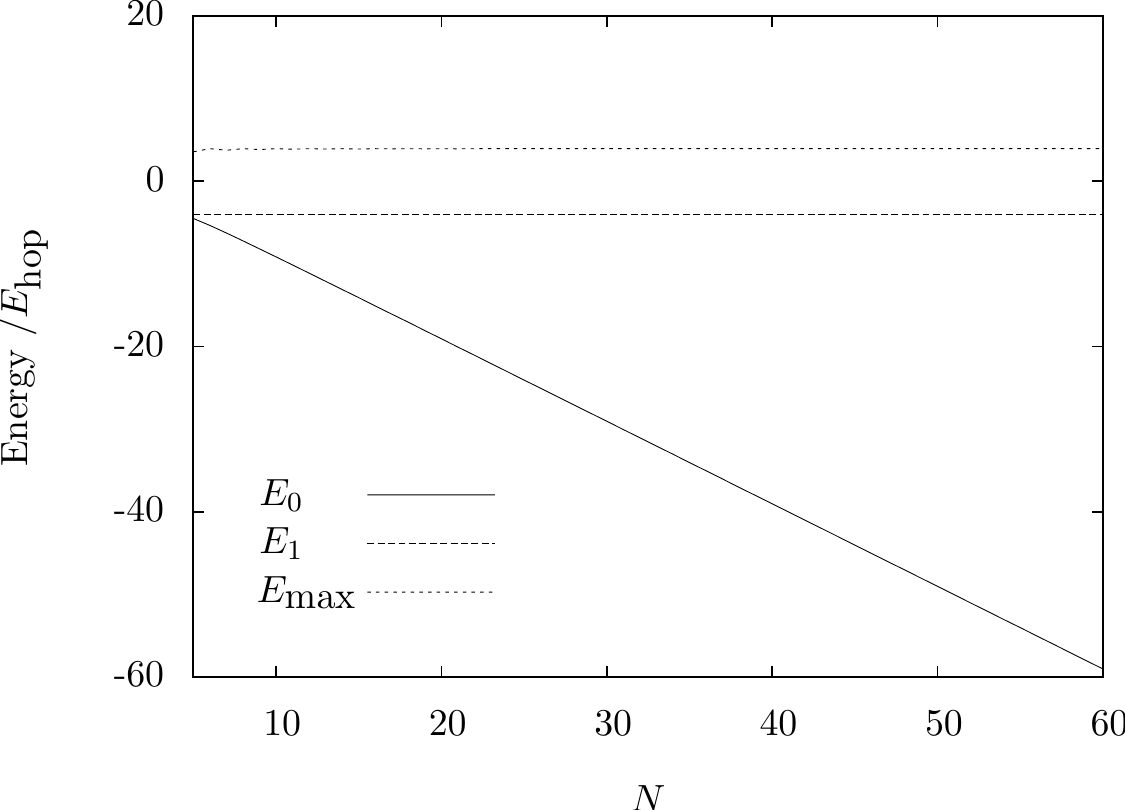}
\caption{Plots of the energy  $E_0$ of the ground state, the energy $E_1$ of the first excited state, and  $E_{\textrm{max}}$, the energy of the  maximum energy state, against
the size  $N$ of the completely connected region of the $\mathscr{K}_N$ graph, in the single particle sector.
This plot has been realized for the full model with $R=30$. Note that the gap $E_1-E_0$ increases linearly with $N$.
}
\label{fig:gap}
\end{figure}

\begin{figure}
\includegraphics[width=\columnwidth]{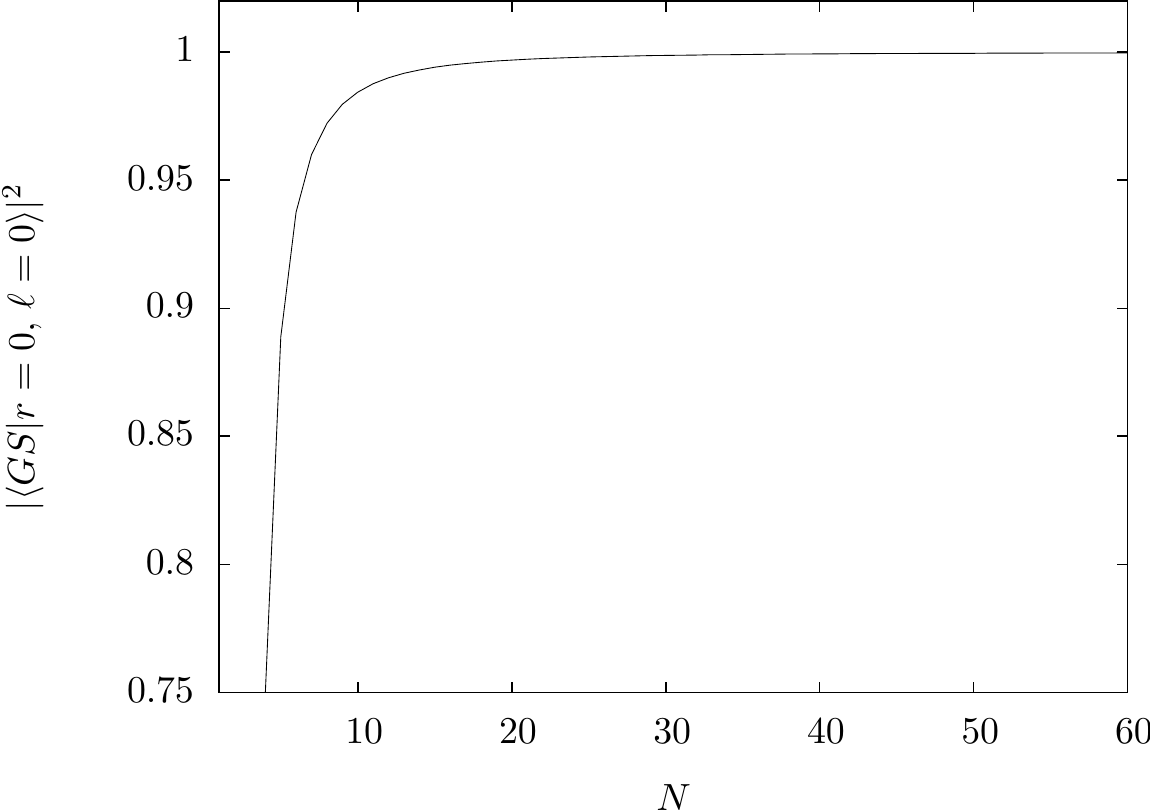}
\caption{Fidelity between the completely delocalized state in the completely connected region $|r=0, \ell=0\rangle$ and the ground state
against the size $N$ of that region. The completely delocalized state $|r=0,\ell=0\rangle$ becomes the ground state of the system for large $N$.
}
\label{fig:fidelity}
\end{figure}

These results are numerically confirmed by Figs.~\ref{fig:gap} and ~\ref{fig:fidelity}. In Fig.~\ref{fig:gap}, the energies of the ground
state, the first excited state and the state with maximum energy are plotted against the size of the completely connected region.
In Fig.~\ref{fig:fidelity}, we have plotted the fidelity between the ground state of the system and the completely delocalized state
in the completely connected region vs $N$. 

%%%%%%%%%%%%%%%%%%%%%%%%%%%%%%%%%%%%%%%%%%%%%%%%%%%%%%

\emph{Multi-particle case}.
Let us next analyze what happens when there are several particles in the $\mathscr K_N$
configuration, interacting with an on-site potential. 

The Hamiltonian of the multi-particle $\mathscr K_N$ model can be decomposed in
its one-body and two-body parts,
\be
\widehat H=\widehat H_0 + \widehat V
\ee
where $\widehat H_0$ is defined in Eq.~\eqref{eq:Hamiltonian-free-Kn} and $\widehat V$ is an on-site interaction among the particles:
\be
\widehat V=u \sum_{r=0}^{R-1}\sum_{\theta=0}^{N-1} b_{r\theta}^\dagger b_{r\theta}^\dagger b_{r\theta}b_{r\theta}\, ,
\ee
with $u$ the energy penalty for two particles in the same site.

Because the interaction $\widehat V$ commutes with the number operator $\widehat N$ and the rotation transformation $\widehat L$,  $\widehat H$, $\widehat L$ and $\widehat N$ form a complete set of commuting observables.
It is convenient to write $\widehat H_0$ and $\widehat V$ in terms of the creation and annihilation operators
\begin{align}
\eta_{k \ell} &=\frac{1}{\sqrt{RN}}\sum_{\theta=0}^{N-1} \sum_{r=0}^{R-1}
	 \e^{\iu \frac{2\pi}{R}k r}e^{\iu \frac{2\pi}{N}\ell \theta}b_{r,\theta}, \ \ \forall \ell \ge 1, \\
\eta_{k 0} &=\frac{1}{\sqrt{RN}}\sum_{\theta=0}^{N-1} \sum_{r=1}^{R-1} \e^{\iu \frac{2\pi}{R-1}r}b_{r,0}, \\
\eta_{\textrm{gs}}&=\frac{1}{\sqrt{N}}\sum_{\theta=0}^{N-1}b_{0, \theta}\, .
\end{align}
As it has been previously shown, the one-body term in the Hamiltonian is
\be
\widehat H_0=-N\eta_{\textrm{gs}}^\dagger\eta_{\textrm{gs}} + \sum_{k\ell}E_{k\ell}\eta_{k\ell}^\dagger\eta_{k\ell}\, ,
\ee
and the interaction reads
\begin{align}
\widehat V&=u\sum_{k \ell} \delta_{\ell_1+\ell_2,\ell_3+\ell_4}\delta_{k_1+k_2, k_3+k_4} 
\widehat\eta_{k_1 \ell_1}^\dagger \eta_{k_2 \ell_2}^\dagger\eta_{k_3 \ell_3}\eta_{k_4 \ell_4}\nonumber \\
&\qquad +u \, \eta_{\textrm{gs}}^\dagger \eta_{\textrm{gs}}^\dagger \eta_{\textrm{gs}}\eta_{\textrm{gs}} \, ,
\end{align}
where the sum runs over all possible $k_i$ and $\ell_i$ for $i=1,\ldots, 4$ and the Kr\"onecker delta
shows that the interaction conserves the quantum numbers $\ell$ and $k$.
Note that the single-particle ground state is completely decoupled from the states of the energy band. 
Therefore, the state of $n$ particles,
\be
\ket{GS}=\left(\eta_{\textrm{gs}}^\dagger\right)^n \ket{0}\, ,
\ee
is an eigenstate of the system with energy $- n N + u n(n-1)/2$.
Furthermore, as the interaction is a positive operator, 
we can ensure that $\ket{GS}$ is the ground state of the system
as long as $n < N$.

Thus, {\em the ground state of the many-body problem is a Bose-Einstein condensate
of delocalized particles at the completely connected region. }
The large gap and the features of the on-site interaction make this condensate
robust at finite temperatures and against adding interacting particles.
Thus, the completely connected region of the $\mathscr K_N$ model can be considered a trapped surface.
%in which the particles trapped cannot escape even if an interaction among them and temperature are introduced.

However, it is not clear whether every completely connected region is a trapped surface.
In particular, we would like to see what happens when the connectivity does not change as abruptly as in the $\mathscr K_N$ system. 
For this reason, in Section \ref{section:falloff} we will parametrize the fall-off of the parameters  and study the localization of
particles.
{Moreover, we would like to relate this fall-off  to an effective curved space-time geometry, as in an analogue gravity system.
 In order to do this, then, a correspondence between the connectivity of the graph and the curvature of the space-time is required. Establishing this relation  is our task for the next section.}

%%%%%%%%%%%%%%%%%%%%%%%%%%%%%%%%%%%%%%%%%%%%%%%%%%%%%%%%%%%%%%%%%%%%
\section{Correspondence between graph connectivity and curved geometry}
\label{section:EOM}

In this section we establish a relation between the connectivity of a graph and
the curvature of its continuous analogue geometry. 
In order to do this, we restrict the time-dependent Schr\"odinger equation  to 
the manifold formed by the classical states, that is, single-particle states with a well-defined but unknown position.
The equation of motion obtained corresponds to the equation of propagation of light in  inhomogeneous media, similarly to black hole analogue systems.
Once we have such a wave equation, we can extract the corresponding metric.  
In the second part of the section,  the dispersion relation and the continuous 
limit are discussed for the transitionally invariant case. 

\subsection{Restriction of the time-dependent Schr\"odinger equation to the set of classical states}

Since we want to study the dynamics of a single particle on a fixed graph, 
it is only necessary to consider the single particle sector.
The one dimensional Bose-Hubbard model for a single particle reads,
\be
\widehat H_0 = \sum_{n=0}^{M-1} f_{n,n+1}\left(\ket{n}\bra{n+1}+\ket{n+1}\bra{n}\right) + \sum_n \mu_n \proj{n}\, , 
\label{eq:Hamiltonian-BH-single-particle}
\ee
where $f_{n,n+1}$ are the tunneling coefficients between sites $n$ and $n+1$,
$\mu_n$ is the chemical potential at the site $n$, {and $M$ is the size of the lattice.

% 
% \begin{figure}
% \includegraphics[scale=0.8]{1d}
% \caption{The multi-link 1-d lattice.}
% \label{1d}
% \end{figure}
%\textcolor{orange}{
% \ar{\sout{
% In this setup, we can define the field position $\Psi_n(t)$ as 
% the probability of finding the particle at time $t$ in the site $n$,
% }
% \be
% \cancel{
% \Psi_n(t):=\bra{n} \rho(t) \ket{n}\, ,}
% \ee
% \sout{
% where $\rho(t)$ the density matrix of the system at time $t$.
% }}

In this setup, let us introduce the convex set of classical states $\mathcal{M}_C$, 
parameterized as
\be
\rho(\Psi) = \sum_{n=0}^{M-1} \Psi_n \proj{n}\, ,
\ee
where $\Psi_n$ is the probability of finding the particle at the site $n$.
The states in $\mathcal{M}_C$ are classical because the uncertainty in the position is classical, that is,
they represent a particle with an unknown but well-defined position.

The aim of this section is to restrict the Schr\"odinger equation of the whole convex set of density matrices to 
the convex set $\mathcal{M}_C$ and obtain the effective 
equations of motion for the classical states. 
In order to do this we will follow the same procedure as in Ref.\ \cite{Haegeman:2011}.
We will approximate the time evolution generated by $\widehat H_0$ without ever leaving the convex set $\mathcal{M}_C$.
This procedure consists basically of two steps:
\begin{enumerate}
\item \emph{Time evolution}. Insert the initial state $\rho(t) \in\mathcal{M}_C$ into the time-dependent Schr\"odinger equation
to get its evolution $\rho(t+\Delta t)$ after a short time $\Delta t$.
\item \emph{Restriction to $\mathcal{M}_C$}. Find the state in $\mathcal{M}_C$ that best approximates the
evolved state $\rho(t+\Delta t)$.
\end{enumerate}
If we take the infinitesimal limit $\Delta t \to 0$ of the previous steps we are going to get a differential equation for the field 
$\Psi_n(t)$.

Let us mention that the time-dependent Gross-Pitaevskii equation can  also be derived by this method, that is, by
restricting the Schr\"odinger equation to the manifold defined by states parametrized by 
$\ket{\varphi}=\exp \left( \int \dd x \varphi (x) b_x^\dagger \right) \ket{0}$, where 
$b_x^\dagger$ is the field operator that creates
a particle at position $x$.
In Ref.\ \cite{Haegeman:2011}, the second step is something that 
must be done in order to restrict the equations of motion to the desired manifold.
Nevertheless, in our case, \emph{decoherence} can give a physical interpretation to this step. 
Since our particle is under the effect of a noisy environment,
its density matrix is going to be constantly dephased by the interaction between the particle and its reservoir.

If we insert $\rho(t)$ in the time-dependent Schr\"odinger equation we obtain:
\be
\partial_t \rho(t)=\sum_n \partial_t \Psi_n(t) \proj{n}=\iu [\widehat H_0, \rho(t)]\, ,
\ee
where, in general, the right hand side cannot be written in terms of the left hand side,
and therefore the previous equation cannot be fulfilled. 
Note that the Hamiltonian pushes  $\rho(t+\Delta t)$ out of the $\mathcal{M}_C$.

Thus, the best approximation to $\rho(t+\Delta t)$ in the convex set $\mathcal{M}_C$ 
is obtained by minimizing the distance
\be
\norm{\sum_n \partial_t \Psi_n(t) \proj{n}-\iu \left[\widehat H_0,\rho(t)\right]}\, ,
\ee
For the Hamiltonian \eqref{eq:Hamiltonian-BH-single-particle}, the commutator $[\widehat H_0,\rho(t)]$ reads
\begin{align}
[\widehat H_0,&  \rho(t)] \\
&=\sum_n f_{n,n+1}\left(\Psi_{n+1}-\Psi_{n}\right)\left(\ket{n}\bra{n+1}+\ket{n+1}\bra{n}\right) \, ,\nonumber
\end{align}
and therefore the state in $\mathcal{M}_C$ that best approximates $\rho(t+\Delta t)$ fulfills
\be
\partial_t \Psi_n(t)=0\, .
\ee
This forces us to consider the second order in $\Delta t$, and 
therefore the effective equation of motion for $\Psi_n(t)$ is going to be
a second order differential equation in time. We then have
\be
\partial_t^2\rho(t)=\sum_n \partial_t^2 \Psi_n(t) \proj{n}=-\frac{1}{\hbar^2}\left[\widehat H_0,[\widehat H_0,\rho(t)]\right] \, .\nonumber
\ee

The dephased state in the position eigenbasis that best approximates $\rho(t+\Delta t)$ can be easily determined by computing
the double commutator of the previous equation. 
It obeys the evolution
\begin{align}
\frac{\hbar^2}{2}\partial_t^2 \Psi_n(t)=& -\left(f_{n,n+1}^2+f_{n-1,n}^2\right)\Psi_n(t) \\
	&\qquad +f_{n-1,n}^2\Psi_{n-1}(t) +f_{n,n+1}^2\Psi_{n-1}(t) \, .\nonumber 
\label{eq:second-derivative-field}
\end{align}
In order to rewrite the previous expression in a nicer way,
let us add and subtract the quantity $f_{n-1,n}^2 (\Psi_{n+1}(t)+ \Psi_{n}(t))$ in the right-hand side, obtaining
\begin{align}
%\label{eq:EOM}
\frac{\hbar^2}{2}\partial_t^2 \Psi_n(t) =& f_{n-1,n}^2 \left(\Psi_{n+1}(t)+\Psi_{n-1}(t)-2 \Psi_{n}(t)\right) \nonumber  \\
&+\left(f_{n+1,n}^2-f_{n-1,n}^2\right)\left( \Psi_{n+1}(t)-\Psi_{n}(t)\right) \, .\nonumber
\end{align}
This equation becomes a wave equation in the continuum,
\begin{equation}
\partial_t ^2 \Psi(x,t)-\partial_x \left(c^2(x)\partial_x \Psi(x,t)\right)=0\, ,
\label{eq:continuous-EOM}
\end{equation}
where 
\begin{equation}
 \frac{1}{c(x)}=\sqrt{ \frac{\hbar^2}{2f^2(x)E_{\textrm{hop}}^2} } = \frac{\hbar}{E_{\textrm{hop}}\sqrt{2f^2(x)}}\, ,
\label{eq:speed-of-light}
\end{equation}
and $\Psi(x,t)$ and $f(x)$ are the continuous limit functions of $\Psi_{n}(t)$ and $f_{n,n-1}$ respectively.

{\em We have shown that the equation of motion of the Bose-Hubbard model, restricted to the convex set
formed by classical states, is the wave equation. }
This is a significant result that establishes a relation between the coupling constants of the Bose-Hubbard model
and the speed of propagation of the fields $\Psi(x,t)$.
Equation (\ref{eq:continuous-EOM}) has the same form as the equation for 
propagation of light in media with a space-dependent refraction index, as is also the case in black hole analogue systems \cite{optic}.

A few comments are in order. In  equation (\ref{eq:continuous-EOM}), {\em the constant of the speed of light is quantized}. It is proportional to the 
inverse of the number of links between the nodes in this simplified model. This constant depends on the 
hopping coupling constant of the hamiltonian, $E_{\textrm{hop}}$. 
%Also, even if we have given mass to the particle, $\mu$, the wave equation would for a \emph{massless} scalar quantity, the contribution would cancel out in the equation. 
Finally, even though equation (\ref{eq:continuous-EOM}) is for a scalar quantity, the analogy with a Klein-Gordon field can not be pushed 
too far. The equation refers to the 1-particle density from which we started from, 
$\Psi_n(t)=\bra{n}\rho(t) \ket{n}$, and so is completely classical. A generalization of equation  (\ref{eq:continuous-EOM}) to the case
of many distinguishable interacting particles is given in detail in the Appendix, where we analyze also the effect of a local and non-local interaction. 
%In particular, we prove that the restriction of the time dependent Schr\"odinger equation to the manifold of classical states in the 
%many particle case gives a set of decoupled differential equations like 

\subsection{Dispersion relation and continuum limit}
\label{section:dispersion}
%\textit{Translational invariance.} 
%\ar{Is this part useful or important for the aim of the paper?}
Let us consider in more detail the translational invariant case in which $f_{n-1,n}=f$ and $\mu_n=\mu$ for all $n$.
In this case, the continuous wave equation \eqref{eq:continuous-EOM} becomes
\be
\partial_t ^2 \Psi(x,t)-c^2\partial_x^2  \Psi(x,t)=0\, ,
\ee
where $c$ is the speed of propagation. 

In order to understand the continuum limit, we evaluate the dispersion relation for the propagation of probability 
in the translationally invariant case. 
Let us introduce a discrete Fourier transform in the spatial coordinate and a continuous Fourier transform in the temporal coordinate, 
given by
\be
\Psi_n(t) = \frac{1}{\sqrt{M}}\sum_{k=0}^{M-1} \tilde \Psi_k(t) e^{-\iu \frac{2\pi}{M} nk}\, ,
\ee 
and $\tilde \Psi_k(t)= A e^{\iu \omega_k t}+ B e^{-\iu \omega_k t}$.
% then the EOM \eqref{eq:EOM} becomes, in the $c=const$ case we have
% \begin{align}
%  & \sum_{k=1}^{N-1}[ (- c^2 \partial_t^2 \tilde \Psi_k(t) + e^{i \frac{2\pi}{N} k}+e^{i \frac{2\pi}{N}k}-2) e^{-i \frac{2\pi}{N} nk}]=\nonumber \\
%  & =\sum_{k=1}^{N-1} [- c^2 \partial_t^2 \tilde \Psi_k(t) + 2(\cos(\frac{2\pi}{N}k)-1))] e^{-i \frac{2\pi}{N} nk},
% \end{align}
After a straightforward calculation, we find that the relation between $\omega_k$ and $k$ is given by
\begin{equation}
\omega_k c = \sqrt{2}\ \sqrt{1-\cos\left(\frac{2\pi}{M}k\right)}.
\label{discdisp}
\end{equation}
Note that this dispersion relation is different from that of the quantum excitations. 
In order to see this, we Fourier transform the Hamiltonian in the translational invariant modes.
We define the field momentum $\Psi_k(t)$ as $\Psi_k(t)=\bra{k}\rho(t)\ket{k}$,
where $\ket{k}=\sum_{m}e^{\iu 2\pi k m / M} /\sqrt{M} \ket{m}$ are the eigenstates
of the translationally invariant Hamiltonian
$\widehat H_0 \ket{k} = \hbar \omega^{H}_k \ket{k}$,
with 
\be
\hbar \omega^{H}_k = 2\left(1-\cos\left(\frac{2\pi k}{M}\right)\right).
\label{dispquant}
\ee
Therefore, 
\be
\Psi_k(t)=\sum_{n,m}e^{\iu 2\pi k (n-m)/M}\rho_{n,m}(t),
\ee
and, for the classical states $\rho_{n,m}(t)=\delta_{n,m} \Psi_{n}(t)$,
\be
\Psi_k(t)=\frac{1}{M}\sum_{n=0}^{M-1}\Psi_{n}(t)=\frac{1}{M}.
\ee
Note now that Eq.~ (\ref{dispquant}) differs from (\ref{discdisp}). In fact, at the quantum level,
the continuum limit gives the ordinary galilean invariance, while at the quantum level, the continuum limit gives excitations 
(time-evolving probability densities) which are Lorentz invariant. 

However, the continuum limit is tricky because we have continuous time and discrete space, and no  \emph{spatial} scale to send to zero with 
the inverse of the number of lattice points (the graph is not embedded in any geometry). 
Thus the continuum  limit is  in the time scale of the modes. 
If we rescale $\omega_k\rightarrow \tilde\omega_k/M$ (or equivalently $c$), we find that 
\begin{equation}
\tilde \omega_k\ c = \sqrt{2M}\sqrt{1-\cos\left(\frac{2\pi}{M}k\right)},
\label{contdisp}
\end{equation}
and, therefore,
$$\lim_{M\rightarrow \infty} \tilde \omega_k\approx 2 \pi \frac{k}{ c}. $$
That is, only  modes that are slow with respect to the time scale set by $c$ see the continuum. Note that by rescaling the speed of propagation $c$, the continuum limit can be obtained by a double scaling limit,
$ E_{\textrm{hop}}\rightarrow E_{\textrm{hop}}/M$ and $M\rightarrow\infty$ for  lattice size $M$. In this limit, the probability 
density has a Lorentz invariant dispersion relation. Another interesting rescaling which gives Lorentz invariant 
dispersion relations is the rescaling of $\hbar$.

To conclude the discussion, let us note that the speed of propagation of the probability sets the timescale of the interaction: 
if there are $P$ points with constant speed of light given by $c$, then the timescale of the propagation of the interaction in the classical limit in that region is given by $t_{\textrm{dyn}}= P/c$. 

An embedding of the graph in space, on the other hand, would mean requiring that $\Psi_n(t)$ depends on a point in space and a length scale $a$ (the lattice spacing):  $\Psi_n(t)\equiv\Psi(b_0+n a ,t)$,  and that the coupling constant scales as
$E_{\textrm{hop}}\rightarrow E_{\textrm{hop}} a$. In this case the continuum limit $a\rightarrow 0$  gives the ordinary dispersion relation.
%\ar{{\bf TO BE COMPLETED:} Discussion of the quantum correlations in the translational invariant case.}

% If the graph is translational invariant, that is,
% $$ f_{n-1,n}=\text{constant} =f,$$
% then the non-closed part of equation (\ref{eq:qeq}), which is the one depending on the non-diagonal part of the density matrix, takes the form
% \begin{equation}
% 2 f^2[\rho_{n+1,n-1}(t)-\rho_{n,n-2}(t) + \rho_{n-1,n+1}(t)-\rho_{n,n+2}(t)]. 
% \label{flat}
% \end{equation}
% Note that this is zero if the $U_{k,k'}$ is translational invariant, that is, if it depends only on $k-k'$. 
% In this case this term vanishes and the equation for the probability evolution is still closed, depends only 
% on $\Psi_n(t)$. It is easy to understand that a completely delocalized excitation, in particular one with a spatial probability distribution
% which is flat, makes Eq.~(\ref{flat}) vanishing. Let consider then an initial 1-particle wavefunction which is gaussian,
% $$|\Psi\rangle=\frac{1}{ \sqrt{ \mathscr N_L}} \sum_{i=0}^L e^{-\frac{(i-i_0)^2}{2\sigma}} |i\rangle, $$ 
% with $\mathscr N_L=\sum_{i=0}^L  e^{-\frac{(i-i_0)^2}{\sigma}}$. In this case we find that
% $$U_{i,i\prime}(0)= e^{-\frac{(i-i_0)^2}{2\sigma}} e^{-\frac{(i\prime-i_0)^2}{2\sigma}},$$
% and then, expanding the exponents assuming $\sigma \gg 1$, after a straightforward calculation we find that this quantity takes the form:\\
% \textcolor{blue}{finish this}

%%%%%%%%%%%%%%%%%%%%%%%%%%%%%%%%%%%%%%%%%%
%%%%%%%%%%%%%%%%%%%%%%%%%%%%%%%%%%%%%%%%%%%%%%%%%%%%%%
\section{Model with parametrized fall-off of connectivity}
\label{section:falloff}
 In this section, we study a graph with a trapped surface (completely connected region) 
 whose boundary is extended and the connectivity of its nodes decreases gradually towards the outer edge of the boundary. 
We parametrize the fall-off of the parameters of the model and study the localization of
a one-particle state. We use the techniques developed in Secs. \ref{section:symmetries} and \ref{section:trapped}
to find the ground state of the system and check its robustness against finite temperature and many interacting particles.

This section summarizes the two main ideas of this paper: the relation between the connectivity of a graph and the optical parameter,
and the analysis of the power of trapping of a completely connected region.

\subsection{The Model}
Let us consider a 2-dimensional rotational invariant graph where the connectivity $f_{r,r+1}$ between two layers $r$ and $r+1$
is not constant ($\mathscr K_N$ case), but decreases as
\be
-f_{r,r+1}=E_{hop} d_{r,r+1}\sqrt{\frac{N_r}{N_{r+1}}}=\left(1+\frac{N}{r^\gamma} \right)\, .
\label{eq:tunneling-coef2}
\ee
$d_{r,r+1}$ is the relative degree of a site of the $r$-th subgraph towards the $r+1$-th subgraph,
$N_r$ is the number of vertices in the layer $r$, and 
$\gamma$ is a parameter that controls how fast the connectivity decreases. 
Note that this choice is consistent because we can always tune $E_{hop}$ to be small enough, {$N_r=N$ uniform, and $d_{r,r+1}$
to be such that we can approximate the rhs of Eq.\ (\ref{eq:tunneling-coef2}).}
The $\mathscr K_N$ model corresponds to $\gamma\to \infty$ together with $N$. We expect that the larger $\gamma$ is, the larger is 
the localization.

% and, using results from
% Sec.~\ref{sec:metric-tunneling}
% the Schwarzschild metric corresponds to $\gamma =1$. 

% Notice that not any connectivity is possible since the relative degree $d_{r,r+1}$ and the number of vertices $N_r$ are integers,
% and the relation $d_{r,r+1}N_r=d_{r+1,r}N_r+1$ must be always fulfilled for a real graph.

\subsection{Trapped surface}

The rotational symmetry allows us to perform the same diagonalization procedure followed in Sec.~\ref{section:trapped}. That is,
we will diagonalize the Hamiltonian in blocks corresponding to the eigenspaces of the rotation operator $\hat L$. 
Unlike the $\mathscr K_N$ graph, the diagonalization in such subspaces must be done numerically.
This is not a problem in the single particle sector. 
We can then determine the ground state and the gap of the system. 

In Fig.~\ref{fig:spectrum-schwarz}, we have found the spectrum of the system. We see that
the ground state is protected by a gap that increases linearly with the size of the completely connected region $N$.
By the same argument as in the $\mathscr K_N$ model, this gap guarantees that 
the completely connected region is a trapped surface robust against finite temperature and against adding several interacting particles.

\begin{figure}
\includegraphics[width=\columnwidth]{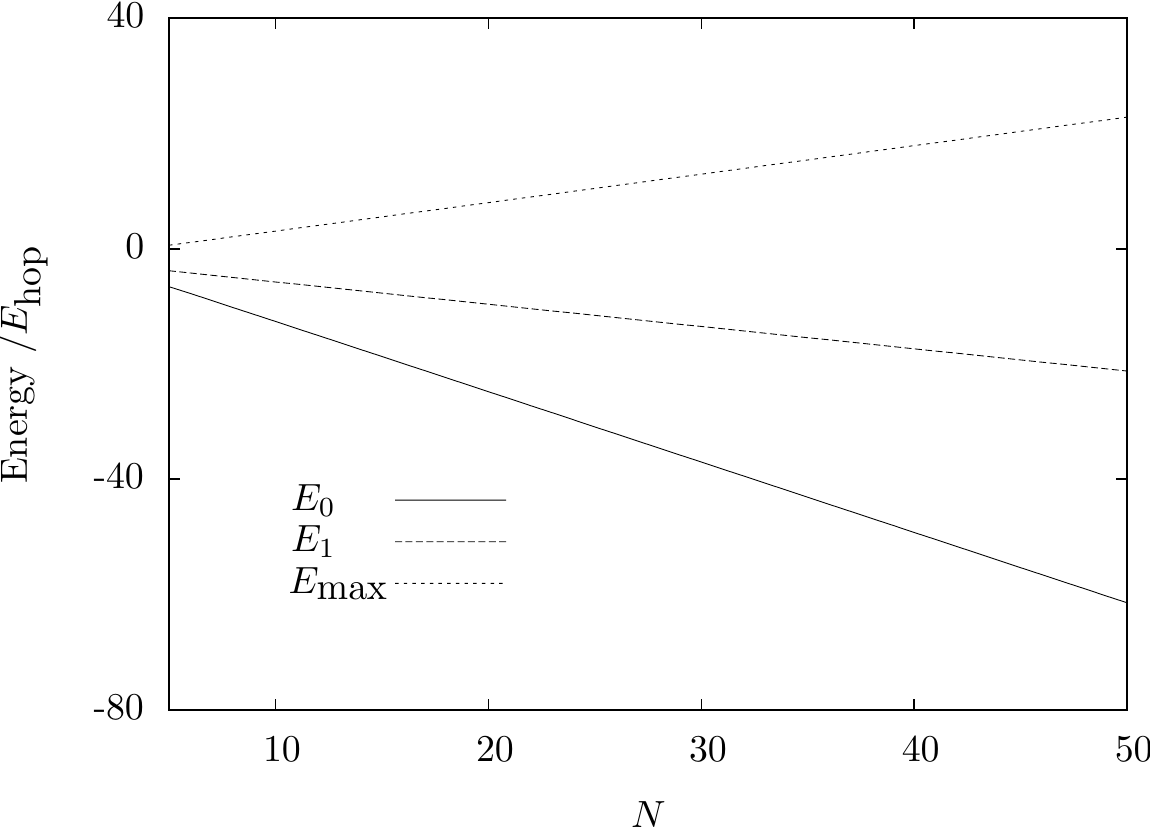}
\caption{
Plot of the energies of the ground state $E_0$, the first excited state $E_1$, and the state with maximum energy $E_{\textrm{max}}$, with 
respect to the size of the trapping surface, $N$, for a rotational invariant graph with decaying connectivity
$(1+N/r)$, in the single particle sector.
Note that the gap $E_1-E_0$ increases linearly with $N$.
}
\label{fig:spectrum-schwarz}
\end{figure}

To characterize the ground state of the system, the probability distribution of the
position of the particle is plot in Fig.~\ref{fig:position-distribution}.
We note that, unlike the $\mathscr K_N$ model, the particle is not completely localized
inside the trapped surface, but localized in a small region in and around the trapping surface.

In Fig.~\ref{fig:particle-localization}, the localization of the particle inside the completely connected region is plotted against  its size and
for three different parameters $\gamma=1/2, 1$ and $2$. 
Both this plot and Fig.~\ref{fig:position-distribution} illustrate that, the higher the parameter $\gamma$, the stronger the localization 
in the completely connected region.

\begin{figure}
\includegraphics[width=\columnwidth]{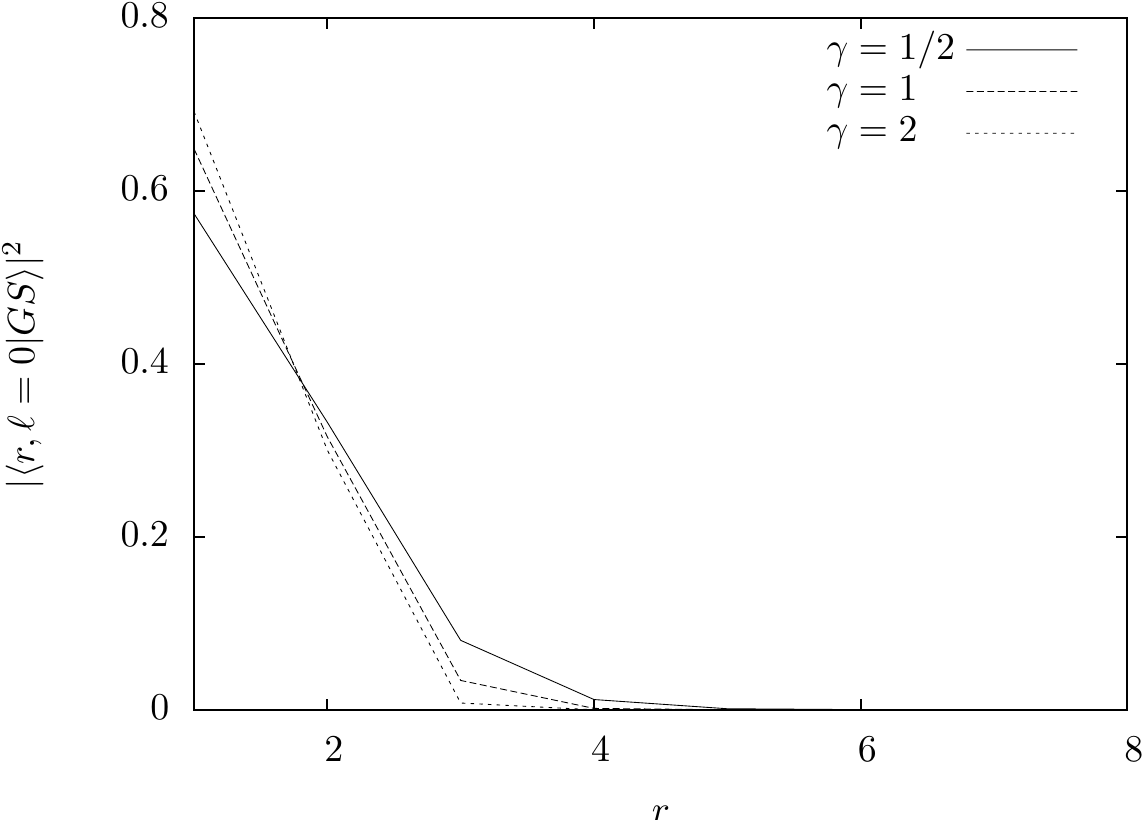}
\caption{
Probability distribution of the position of the particle.
The particle is completely confined around the trapping surface.
The larger the fall-off coefficient $\gamma$, the larger the confinement.
}
\label{fig:position-distribution}
\end{figure}

\begin{figure}
\includegraphics[width=\columnwidth]{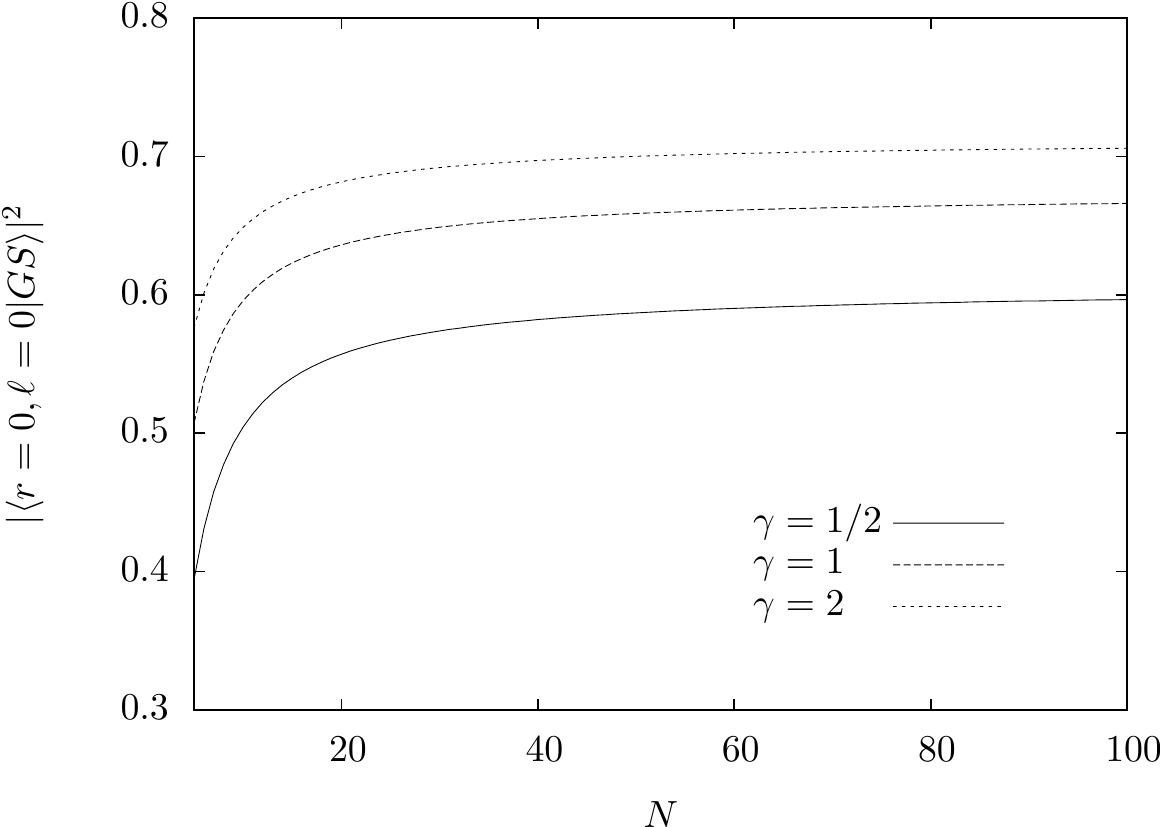}
\caption{
Localization of a single particle in a trapped surface vs its size.
The localization of the particle increases with the size $N$ and
the fall-off coefficient $\gamma$ (see the definition of $\gamma$ in Eq.\ (\ref{eq:tunneling-coef2})).
}
\label{fig:particle-localization}
\end{figure}

%%%%%%%%%%%%%%%%%%%%%%%%%%%%%%%%%%%%%%%%%%%%%%%%

\section{Conclusions}

Quantum mechanics drastically deviates from the ordinary intuition because of its nonlocality, or action-at-distance. Wes do, however, know that 
in condensed matter systems we observe excitations with relativistic dispersion relations. Why is this?  An analogue of this problem is sound in classical systems. 
Newtonian mechanics is nonlocal but Newton himself evaluated the speed of propagation of sound in gases.  

It is now fairly well-understood that theories with nonlocal action but local interactions have  excitations that are locally covariant in 
certain limits. The amount of nonlocality can be bounded using the Lieb-Robinson bounds \cite{liebrob1}.
In this paper we have explored this phenomenon in the Hubbard model.  

We found that the vertex degree plays a central role as it is related to 
the local speed of propagation of the classical probability distribution of matter on the lattice.  
Our analysis is relevant for the idea that relativity, general and special, may be emergent from an underlying theory which is local and 
quantum. Going further than  the analysis of the Lieb-Robinson bounds in \cite{HMPSS},  
we looked at emergence of an effective curved geometry in the Hubbard model.  
The quantum Hamiltonian 
evolution of the one-particle density for a particular set of states leads to an equation of motion that approximates the wave equation for
a scalar particle in curved space. The same equation appears in optical systems with a varying refraction index. 

It is widely known that this 
equation can be cast in the domain of general relativity by noticing that it is the same as evolution in Gordon's space-time. 
This connection cannot be made in 1+1-dimensions,  the case we considered, but equation (\ref{eq:continuous-EOM}) can be extended to
more than one spatial dimension by considering graphs which can be foliated in more than one directions. 
 
Note also that this has been achieved on a Bose-Hubbard model on graphs with varying vertex degree  and multiple edges between sites.  
The important picture to keep in mind to understand this mechanism is Fig.~\ref{fig:scheme}. 
The graph modifies the strength of the 
interaction of the original Bose-Hubbard model and, for the particular states we are interested in,  we can obtain Eq.~(\ref{eq:continuous-EOM}) for the particle 
localization.

Using similar methods, we confirmed
 previous arguments \cite{graphity2} and showed
 that regions of high connectivity in the lattice trap matter. 
 We did this by showing that this ground state is protected by a gap which increases linearly with the size of the completely connected subgraph.  
 
 As with the emergent wave equation, this result is reminiscent of the physics of analogue gravity systems .  It is then interesting to consider the model of \cite{graphity2} as a rather unusal analogue gravity model.  
 
 We will close with the two obvious next questions, which we leave for future work: 1) What is the timescale of thermalization of matter in the completely connected regions?  Does it support the conjecture in 
 \cite{scramble1,scramble2} that black holes are the fastest thermalizers in nature?  2) How do our results change if we include the backreaction of the matter on the graph?  {(Our study makes it essential in order to have evaporation.)} What is the effective dynamics of the full Hamiltonian?

\begin{center}
{\bf Aknowledgements}\\ 
\end{center}
We would like to thank Abhay Ashtekar and John Preskill for initial discussions that motivated this work, and 
Lorenzo Sindoni for several discussions on the relation of the model introduced in the present paper with optical systems.

Research at Perimeter Institute is supported by the Government of Canada
through Industry Canada and by the Province of Ontario
through the Ministry of Research \& Innovation.

%%%%%%%%%%%%%%%%%%%%%%%%%%%%%%%%%%%%%%%%%%%%%%%%
\section*{Appendix: Evolution in the multi-particle sector}
\label{section:Appendix}

In this Appendix, we   generalize the  results of Section \ref{section:EOM} to the case of multiple particles hopping on the graph.

\emph{Free motion.} Let us first consider  $K$ distinguishable particles
in  states which are tensor products of  delocalized states. This will allow us to  track every single 
particle using a number operator.

Since the particles are distinguishable, we cannot use the standard second quantization formalism.  Instead, 
a quantum state of the system must be described by 
\be
	\ket{\Phi_K}=\sum_{m_1,\ldots, m_K = 1}^{\tilde N}c_{m_1,\ldots, m_K }|m_1\rangle \otimes \cdots \otimes |m_K \rangle \, , \nonumber
\ee
where $\ket{m_i}$ is the state of the $i$-th particle at the site $m_i$, and $\widetilde N$ is the total number of sites of the lattice.
Note that the dimension of the Hilbert space is $\widetilde N^K$. 

Let us also introduce a number operators $\widehat N_i $, defined as the tensor product
of the number operator on the site of the particle $i$ with identity operators on all sites except $i$ so that the total number operator is 
 $\widehat N = \sum_{i=1}^{K} \widehat N_i$. 
This number operator counts the total number of particles on the graph. 
To see how to go from the 1-particle case to this number operator, let
 us define $\widehat N^{k}_{i}$ as 
the number operator on the vertex $k$ for the particle $i$. In this case we can write $\widehat N = \sum_{i,k} \widehat N_i^k,$
and define a number operator for a region $A$ of the graph as $\widehat N_A = \sum_{k\in A, i} \widehat N_i^k $.
This consistently measures the number of particles in a region $A$ of the graph. 

It is easiest to start with the 2-particle case and then extend the analysis to $K$ particles.  The main complication compared to the 1-particle case is that we now have a density matrix with $4$ indices.   In the non-diagonal case, we have 
\be
\rho = \sum_{k_1,k_2,k_1\prime,k_2\prime} U_{k_1,k_2,k_1\prime,k_2\prime}(t) 
|k_1\rangle\langle k_2| \otimes |k_1^\prime\rangle\langle k_2^\prime|.
\ee
Fortunately, we also have $\widehat H = \widehat H_1 + \widehat H_2$, with $[\widehat H_1,\widehat H_2]=0$.  
A straightforward calculation shows that we can use the result from the 1-particle case:  
\be
\left[\widehat H,[\widehat H,\rho]\right]=\left[\widehat H_1,[\widehat H_1,\rho]\right]+\left[\widehat H_2,[\widehat H_2,\rho]\right]+\widehat R,
\ee
 where the mixed term is 
\begin{equation}
\widehat R=2 \left(\widehat H_1 \widehat H_2 \rho + \rho \widehat H_1 \widehat H_2-\widehat H_1 \rho \widehat H_2- \widehat H_2 \rho \widehat H_1\right).
\end{equation}
We follow the same steps as in  the 1-particle case and evaluate the second derivative of the expectation value of the two number operators to find
\begin{align}
\partial_t^2 \alpha^1_{k}+ \partial_t^2 \alpha^2_{k} =&-\frac{1}{\hbar^2}\Big( \text{Tr}\left[ [\widehat H_1,[\widehat H_1,\rho]] \widehat N^1_k\right] \nonumber \\
&+\text{Tr}\left[ [\widehat H_2,[\widehat H_2,\rho]] \widehat N^2_k\right]
\nonumber \\
&+\text{Tr}\left[ \widehat R \rho (\widehat N^1_k+\widehat N^2_k)\right] \Big).
\label{eqn:2p}
\end{align}
It is easy to understand what happens 
when $\rho = \rho_1 \otimes \rho_2$, for which $U_{k_1,k_2,k_1\prime,k_2\prime}(t)=U_{k_1,k_2}(t)\tilde U_{k_1\prime,k_2\prime}(t)$:
 the first and second terms on the r.h.s.\ of Eq.~ (\ref{eqn:2p}) reduce to  discrete second-order derivatives, as in the 1-particle case.
We therefore only have to deal  with the $\text{Tr}\big[ \widehat R \rho (\widehat N^1_k+\widehat N^2_k)\big]$ term.

Another way to see this is by noticing that $\widehat N_j$ acts as a projector on the one-particle states. We have
$ \text{Tr}[\widehat H_1 \rho \widehat H_2]
= \text{Tr}[ \widehat H_2 \widehat H_1 \rho ], $
using the properties of the trace. Term by term, we can show that, for each $\widehat H_i$ and $\widehat H_j$, the mixed term vanishes for any number of particles.
Defining $\widehat R_{ij}=\widehat H_i \widehat H_j \rho + \rho \widehat H_i \widehat H_j - \widehat H_i \rho \widehat H_j-\widehat H_j \rho \widehat H_i$, 
we find that the many-particle equation becomes,
\be
\sum_{i} \partial_t ^2 \alpha^i_k(t) = \sum_i {\nabla\alpha^i_k}+ 2 \text{Tr}\left[\left[\sum_{<ij>}\widehat R_{ij}\right] \sum_i \widehat N^i_k\right],
\ee
where $\nabla \alpha^i_k$ is the second order discrete derivative $\partial_x (c\partial_x (\cdot))$. 

We can now  use the same argument as in the 2-particle case
to show that all the terms $\widehat R_{ij}$ vanish independently. We define $\alpha_k:=\frac{1}{2}\sum_i \alpha^i_k$ to be  the probability of finding a particle at $k$. 
This quantity satisfies the same equation as the one-particle sector probability:
\begin{equation}
 \sum_{i} \partial_t ^2 \alpha^i_k(t) = \sum_i {\nabla\alpha^i_k}.
 \label{eq:multpar}
\end{equation}
That is, particles are independent from each other and each follows its own equation (\ref{eq:multpar}).\\

\emph{Interaction.} 
Including interaction, in general, the Hamiltonian of the system can be written as
\be
\widehat H= \sum_{i=1}^K \widehat H_i + \sum_{i<j} \widehat H_{ij} \, ,
\ee
where 
$\widehat H_i= \underbrace{\hat 1 \otimes \cdots \hat 1}_{i-1\ \text{times}}
 \otimes \widehat H \otimes \underbrace{\hat 1 \otimes \cdots \otimes \hat 1}_{K-i-1\ \text{times}}$. %}_{K-i-1\ \text{times} }$.   
For what follows, again we first check the case of two particles,  and calculate what happens to  Eq.~ (\ref{eq:multpar})  when we add an interaction term. The most general interaction Hamiltonian is of the form
$\widehat H_{int} = \sum_{ijkl} \widetilde U_{ijkl}\ \widehat a^{1 \dagger}_i  \widehat a^{1}_j \otimes \widehat a^{2 \dagger}_k \widehat a^{2}_l$.
We make a simplifying assumption which is natural for delocalized states, and require that the potential respects their symmetries.  
The simplest such local potential is of the form
\begin{equation}
\widehat H_{int} = \sum_{k} I_{k} |k\rangle\langle k| \otimes |k\rangle\langle k|. 
\label{eq:int2}
\end{equation}
If we add this hamiltonian to the free one, additional terms appear on the rhs in Eq.~ (\ref{eq:multpar}). 
These  can be traced back to the additional commutators in the expansion of the full Hamiltonian,
\begin{align}
\widehat C :=& \left[\widehat H_1,[\widehat H_{int},\rho]\right] +\left[\widehat H_{int} ,[\widehat H_1,\rho]\right] \nonumber \\
           &+(1\rightarrow 2)+  \left[\widehat H_{int}, [\widehat H_{int} ,\rho]\right],
\label{eq:commint}
\end{align}
that come from expanding the full hamiltonian $\widehat H_T=\widehat H_1+\widehat H_2 + \widehat H_{int}$ in the double commutators and keeping only the terms involving free Hamiltonians. 
Since we want to distinguish the single particles in the continuum limit, we consider the approximation  $\|\widehat H_{int}\|\ll\|\widehat H_{1/2}\|$, with $\| \widehat T \| = \text{sup}_{v} \frac{\| \widehat T v\|}{|v|} $.  This gives a product of the density matrices of the single particles.
  
A lengthly but straightforward calculation shows that, when we restrict to diagonal density matrices in the subspace of the single particles, we find:
$$\text{Tr}[\widehat C \widehat N^1_k]=\text{Tr}[\widehat C \widehat N^2_k]=0. $$
This means that, surprisingly, in the continuum limit, the wave equations for the fields are decoupled if we use a local potential of the form (\ref{eq:int2}). 

The next generalization  is a potential which is slightly non-local. The simplest such potential is given by
\begin{equation}
 \widehat H_{int}=\sum_k I_k\ |k+1\rangle\langle k+1| \otimes |k\rangle\langle k|
\label{eq:intnl}
\end{equation}
for the case of 2-particles. In this case, an even longer but still straightforward evaluation of $\text{Tr}[\widehat C N^1_k]$ or $\text{Tr}[\widehat C N^2_k]$ shows that these traces are nonzero.  They take the form
\begin{equation}
 \text{Tr}[\widehat C N^1_k]= f_{k,k} U_{k,k}(t)\tilde U_{k+1,k+1}(t) (I_{k+1}-I_{k}),
\label{eq:inttrace1}
\end{equation}
and
\begin{equation}
 \text{Tr}[\widehat C N^2_k]= f_{k,k} U_{k+1,k+1}(t)\tilde U_{k,k}(t) (I_{k+1}-I_{k}).
\label{eq:inttrace2}
\end{equation}
That is, in the continuum limit, the equations for the two probability fields are coupled. If we define  $\alpha_k(t):=U_{k,k}(t)$ and $\beta_k(t):=\tilde U_{k,k}$ the probability fields obey
\begin{align}
& \partial_t^2 \alpha_k-\nabla \alpha_k = \alpha_k \beta_{k+1} f_{k,k} (I_{k+1}-I_{k}) \\
& \partial_t^2 \beta_k-\nabla \beta_k = \beta_k \alpha_{k+1} f_{k,k} (I_{k+1}-I_{k})
\end{align}
If we define $\mu(x)$ to be the continuum equivalent of $f_{k,k}$ and $I(x)$  the continuum equivalent of $I_k$, we have
\begin{align}
& \Box \alpha(x,t) = \alpha(x,t) \beta(x,t) \mu(x)  I^\prime(x) \\
& \Box \beta(x,t) = \beta(x,t) \alpha(x,t) \mu(x) I^\prime(x).
\end{align}
These equations can be straightforwardly generalized to the case of more than two particles if the potential is the sum of a 2-body interaction for each pair of particles of the type
$ \sum_{\langle i j \rangle} \widehat H_{int}^{ij} $,
with 
\begin{align}
\widehat H_{int}^{ij}= \sum_{k}  \hat 1 &\otimes \cdots \hat 1 \otimes \underbrace{|k\rangle\langle k|}_{i-\text{th}} \nonumber \\
&\otimes \hat 1 \otimes \cdots\otimes \hat 1  \otimes \underbrace{|k-1\rangle\langle k-1|}_{j-\text{th}} \otimes \hat 1 \cdots \otimes \hat 1\ I^{ij}_k.
\label{eq:interactionmb}
\end{align}
The sum $\langle i j \rangle$ is over all pairs of particles. If  $\alpha^i(x,t)$ is the probability field  for particle $i$, we obtain, 
for each particle, the following coupled differential equations:
\begin{equation}
 \Box \alpha^i = \alpha^i (\sum_{j\neq i} {\partial I}^{ij} \alpha^j) \mu.
\end{equation}
We can now define $ m_i^2=(\sum_{j\neq i} {\partial I}^{ij} \alpha^j) \mu$. The EOM becomes,
$$ \Box \alpha^i - m_i ^2 (x,t) \alpha^i =0 $$
The effective mass $m_i^2$ is determined by the interaction with the other particles. 
The study of the interaction is important to understand, but beyond of the scope of the paper.

%%%%%%%%%%%%%%%%%%%%%%%%%%%%%%%%%%%%%%%%%%%%%%%%%

\end{document}